\documentclass[pra,twocolumn,superscriptaddress]{revtex4-2}

\usepackage{graphicx}
\usepackage{amsmath,amssymb}
\usepackage{natbib}

\usepackage[colorlinks,linkcolor=blue,anchorcolor=blue,urlcolor=blue,citecolor=blue]{hyperref}
\usepackage[all]{hypcap}

\usepackage{mathrsfs}
\usepackage{bm}
\usepackage{bbm}

\usepackage{dsfont}

\usepackage[utf8]{inputenc}
\usepackage[english]{babel}

\usepackage[T1]{fontenc}

\usepackage{newtxtext}
\usepackage{newtxmath}

\newcommand{\ket}[1]{\left\vert #1 \right\rangle}
\newcommand{\bra}[1]{\left\langle #1 \right\vert}
\newcommand{\ketbra}[2]{\left| #1\right>\left< #2 \right|}
\newcommand{\braket}[2]{\left< #1| #2 \right>}

\begin{document}
\title{Simulating superluminal propagation of Dirac particles using trapped ions}
\author{Qianqian Chen}
\author{Yaoming Chu}
\email{yaomingchu@hust.edu.cn}
\affiliation{School of Physics, Hubei Key Laboratory of Gravitation and Quantum Physics, Institute for Quantum Science and Engineering, International Joint Laboratory on Quantum Sensing and Quantum Metrology, Huazhong University of Science and Technology, Wuhan 430074, China}
\author{Jianming Cai}
\email{jianmingcai@hust.edu.cn}
\affiliation{School of Physics, Hubei Key Laboratory of Gravitation and Quantum Physics, Institute for Quantum Science and Engineering, International Joint Laboratory on Quantum Sensing and Quantum Metrology, Huazhong University of Science and Technology, Wuhan 430074, China}
\affiliation{Wuhan National Laboratory for Optoelectronics, Huazhong University of Science and Technology, Wuhan 430074, China}
\affiliation{State Key Laboratory of Precision Spectroscopy, East China Normal University, Shanghai, 200062, China}
\date{\today}
\begin{abstract}
Simulating quantum phenomena in extreme spacetimes in the laboratory represents a powerful approach to explore fundamental physics in the interplay of quantum field theory and general relativity. Here we propose to simulate the movement of a Dirac particle propagating with a superluminal velocity caused by the emergent Alcubierre warp drive spacetime using trapped ions. We demonstrate that the platform allows observing the tilted lightcone that manifests as a superluminal velocity, which is in agreement with the prediction of general relativity. Furthermore, the Zitterbewegung effect arising from relativistic quantum mechanics persists with the superluminal propagation and is experimentally measurable. The present scheme can be extended to simulate the Dirac equation in other exotic curved spacetimes, thus provides a versatile tool to gain insights into the fundamental limit of these extreme spacetimes.
\end{abstract}
\maketitle

%\tableofcontents
%\newpage
\section{Introduction}
Quantum field theory in curved spacetime \cite{Davis1976,Jacobson2005,Hollands2015}, as a semi-classical approach to study the movement of quantized matter fields in a fixed gravitational background, has predicted many striking phenomena such as the famous Hawking radiation and cosmological particle production. However, confirmation of those extremely weak quantum effects in real gravity experiments remains highly unlikely with present technology  \cite{Howl2018,Howl2021}. Following Unruh's pioneering work \cite{Unruh1981}, much attention has turned into sound waves travelling against the classical \cite{Weinfurtner2011} and quantum fluids \cite{Zurek1985,Fedichev2004,Garay2000,Barcelo2003,Fedichev2003,Fischer2004,Carusotto2008,steinhauer2014observation,Steinhauer2016,Cha2017,De2019,Isoard2020}, or light in nonlinear optical platforms  \cite{Belgiorno2010,Philbin2008,Drori2019,Lang2019} to effectively emulate quantum field theory in curved spacetime. Remarkably, these relevant analogue experiments among many others \cite{,Boiron2015,Roldan-Molina2017} has brought us a better understanding of the relationship between space-time structure and quantum theory  \cite{Chuang1991,Barcelo2011,Faccio2013}.
Furthermore, recent developments of precise and flexible quantum control also facilitate a quantum simulation of related relativistic phenomena \cite{Alsing2005,Schutzhold2007,Nicolas2010,Casanova2010,Gerritsma2011,Wittemer2019,TianZehua2020,YangRunQiu2020,Nation2009,Sabin2016,TianZehua2017,TianZehua2019}. As a representative example, intriguing quantum dynamics related to a Dirac particle moving in curved spacetime has attracted increasingly investigations via the particularly promising trapped-ion platform \cite{Collas2019,Pedernales2018,Garcia2019,Lamata2007}. The exceptional controllability of quantum simulators makes it feasible to emulate a Dirac particle in "exotic" curved spacetimes \cite{Garcia2019}. As a particularly important example, the Alcubierre warp-drive, consistent with Einstein's field equations, can result in faster-than-light propulsion of a spaceship \cite{alcubierre1994}. Compared to the extreme difficulty in realizing such a time-machine model (i.e. creating warp bubbles) in the actual world \cite{Finazzi2009,White2003,White2011}, observation of analogue phenomenon with well-controllable quantum simulators in the laboratory should be more accessible. Besides, merging of fundamental concepts from different fields when implementing such novel quantum simulation, including gravitation, quantum squeezing and quantum entanglement, would provide a fruitful way to reveal unique features of quantum effects in both the exotic curved spacetimes \cite{Garcia2019} and the basic light-matter interaction models \cite{Pedernales2018}.
In this Letter, we propose a trapped-ion quantum simulation of a Dirac particle propagating with a superluminal velocity caused by the emergent Alcubierre warp drive spacetime \cite{alcubierre1994}. The Hamiltonian from the Dirac equation in the Alcubierre (1+1)-dimensional universe is mapped onto a spin-boson interaction quantum model, which can be further realized by a combination of sideband drives and a periodic modulation of the trapping potential. Remarkably, the tunability of system parameters in the well-controlled trapped-ion platform allows us to access the crossover from flat to curved spacetimes. Using exact numerical simulations, we demonstrate that this platform is able to observe analogue superluminal travel of Dirac particles in the tilted light cones, as well as the Zitterbewegung effect of massive Dirac particles incorporating the tilt of the Dirac cone. The extension of the present scheme to simulate the Dirac equation in more general exotic curved spacetimes would make it possible to explore intriguing phenomena arising from the interplay of quantum field theory and general relativity in well-controllable quantum experiments in the laboratory. 
\section{Simulation of Alcubierre metric using trapped ions}
We start by first introducing the Alcubierre metric \cite{alcubierre1994,Gonzalez-Diaz2000}, which is given by
\begin{equation}
  ds^2=c^2dt^2-\left[dx-v_s(t)dt\right]^2-dy^2-dz^2,
\end{equation}
where $v_s(t)\equiv dx_s(t)/dt$ is the velocity associated with a certain trajectory $x_s(t)$. In the (1+1)-dimensional case, the light cones at a point in the $t$-$x$ plane are specified by the curves emerging from the point with $ds^2 = 0$, namely
\begin{equation}\label{eq:velocityGR}
 \frac{1}{c} \frac{dx}{dt}=v_s(t)\pm1.
\end{equation}
If $v_s(t)=0$, the the spacetime is flat. Otherwise the corresponding light cones are tipped over and the particle travelling inside the light cone can have a velocity faster than the light speed in the flat spacetime, which is theoretically consistent with the framework of general relativity \cite{alcubierre1994}.
Extension of the Dirac equation into curved spacetimes successfully merges quantum mechanics with the general relativity. In the (1+1)-dimensional spacetime with signature $(+,-)$, the Dirac equation reads \cite{Mann1991,Collas2019}
\begin{equation}\label{eq:DE}
  \left(i\hbar\gamma^a e^\mu_{\,(a)}\partial_\mu+\frac{i\hbar}{2}\gamma^a\frac{1}{\sqrt{-g}}  \partial_\mu\left(\sqrt{-g}e^\mu_{\,(a)}\right)-mc\right)\psi=0,
\end{equation}
where the Greek letter $\mu$ and the Latin letter $a$ denote the coordinates of curved spacetime and local rest frame respectively, $m$ is the mass of the quantized field, $g$ is the determinant of the metric tensor, and $e^\mu_{\,(a)}$ is the vielbein allowing the constant Dirac matrices $\gamma^a$ to act at each spacetime point. Here, we choose the chiral representation such that $\gamma^0=\sigma_x$, $\gamma^1=-i\sigma_y$, with  ${\bm \sigma}$ the Pauli matrices. In the specific case of Alcubierre metric, the Dirac equation can be  transformed into the Schr{\"o}dinger equation of the following form \cite{Supplement}
\begin{figure}[t]
  \centering
  \includegraphics[width=8.6cm]{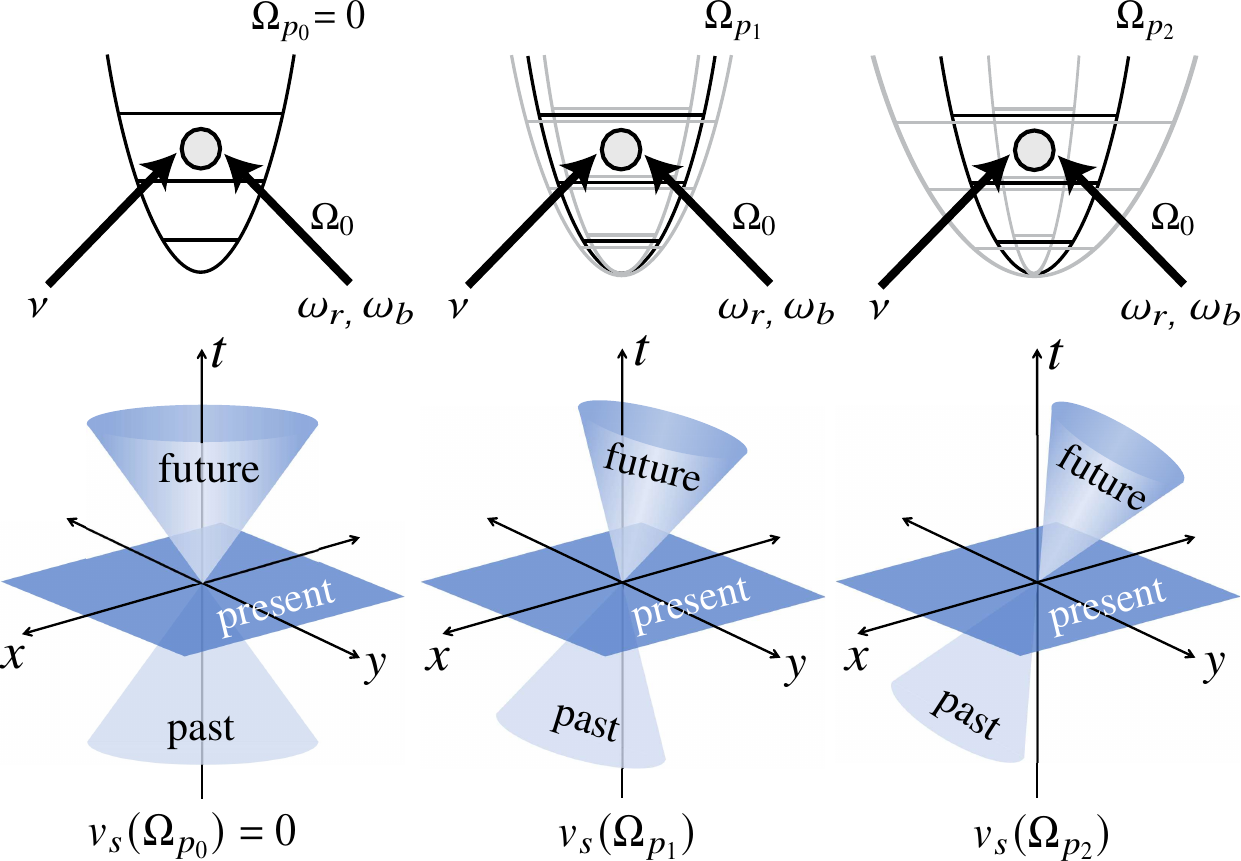}
  \caption{Schematic illustration of the light cones for Dirac particles in the Alcubierre warp drive spacetime which can be simulated by a trapped-ion system. The key ingredients include: (I) the qubit-motion coupling $\Omega_0$ which is implemented by blue and red sideband drives at frequencies $\omega_b$ and $\omega_r$ respectively; (II) the modulation of the confining potential on resonance with the ion-trapping frequency $\nu$, the amplitude of which is given by the parametric coupling strength $\Omega_p$. The tilt of the light cones increases as the parametric coupling strength $\Omega_p$ and thereby the Alcubierre metric parameter $v_s$ becomes larger, which results in a superluminal velocity. The tunability of the control parameters within the state-of-the-art experiments makes it feasible to simulate the tilted Dirac equation in the Alcubierre metric in different parameter regions.}
  \label{fig:tilt_lightcones}
\end{figure}
\begin{equation}\label{}
  i\hbar\partial_t\psi=\left(-i c\hbar v_s(t)\frac{\partial}{\partial x}-ic\hbar\sigma_z\frac{\partial}{\partial x}-mc^2\sigma_x\right)\psi,
\end{equation}
from which the Hamiltonian $H$ that governs the above dynamical evolution can be written as
\begin{equation}\label{eq:H}
  H=cA\hat{P}\sigma_z+cA\hat{P} v_s(t)-mc^2\sigma_x,
\end{equation}
Here we introduce the operators $\hat{X}$ and $\hat{P}$
\begin{equation}\label{eq:XP}
  \hat{X}\equiv A \hat{x},\quad   \hat{P}\equiv-i\hbar\frac{\partial}{\partial X}
\end{equation}
to rescale the spatial coordinate $x$ of the simulated Dirac particle by a dimensionless factor $A$. Note that $\hat{X}$ and $\hat{P}$ will be the observables of the trapped ion in our proposed simulation platform. We remark that the second term of the Hamiltonian in Eq.\eqref{eq:H}, which is linearly dependent on the momentum $\hat{P}$, is an evidence of relativistic physics \cite{thaller2013}. Based on the standard commutation relation $[\hat{X},\hat{P}]=i\hbar$, the operators $\hat{X}$ and $\hat{P}$ can be further mapped to a bosonic field of frequency $\nu$
\begin{equation}\label{eq:bo_field}
  \hat{X}=\sqrt{\frac{\hbar}{2m_0\nu}}(\hat{a}+\hat{a}^\dagger), \quad \hat{P}=-i\sqrt{\frac{\hbar m_0\nu}{2}}(\hat{a}-\hat{a}^\dagger),
\end{equation}
where $m_0$ is a constant with the dimension of mass.
By substituting Eq.\eqref{eq:bo_field} into Eq. \eqref{eq:H} and applying the Hadamard transformation (i.e. $\sigma_x\to\sigma_z$, $\sigma_z\to \sigma_x$), the Hamiltonian can be rewritten as
\begin{equation}\label{eq:Haa}
H=-icA\sqrt{\frac{\hbar m_0\nu}{2}}(a-a^{\dagger})[\sigma_{x}+v_s(t)]-m c^{2} \sigma_{z}.
\end{equation}
To simulate the above Hamiltonian in Eq.\eqref{eq:Haa} with the tunable parameter $v_s$, we consider a setup of a single ion trapped above a linear surface electrode radio-frequency trap. The radial motional mode of the trapped ion with a frequency $\nu$ models a harmonic oscillator and its two internal ground hyperfine levels with the transition frequency $\omega_0$ plays the role of an effective spin-1/2, denoted as $|{\uparrow}\rangle$ and $|{\downarrow}\rangle$. As an example, the present scheme may employ a trapped ${}^{25}$Mg$^+$ ion hyperfine qubit, with an out-of-phase radial motional mode frequency $\nu\approx2\pi\times5.9$ MHz \cite{Burd2021}. The qubit states are chosen as $\left|\downarrow\right\rangle \equiv\left|F=3, m_{F}=1\right\rangle$ and $\left|\uparrow\right\rangle \equiv\left|F=2, m_{F}=1\right\rangle$ \cite{Burd2021,Burd2019}, where $F$ is the total angular momentum and $m_F$ is its projection along the quantization axis.

The spin-harmonic oscillator coupling of the form in Eq.\eqref{eq:Haa} can be realized by implementing the Mølmer$-$Sørensen interaction \cite{Sorensen1999} via simultaneous blue and red sideband drives at frequencies $\omega_b$ and $\omega_r$ respectively, which in real experiments can be generated by using oscillating near-field magnetic field gradients \cite{Srinivas2019}. In addition, we introduce a periodic modulation of the single ion’s trapping potential of the strength $\Omega_p(t)$ at frequency $\nu$ to obtain the second term $\sim cv_s(t)\hat{P}$ in Eq.\eqref{eq:Haa}. This can be experimentally implemented by applying an oscillating potential directly to the radio-frequency trapping electrodes, as demonstrated in Ref.\cite{Burd2021}.
Such a trapped-ion platform can be characterised by the following system Hamiltonian as \cite{Supplement}
\begin{eqnarray}\label{eq:H_lab}
H=&&\frac{\hbar\omega_0}{2}\sigma_z+\hbar\nu a^\dagger a +\hbar\Omega_p(t)\sin(\nu t)(a+a^\dagger)^2 \\\nonumber
&&+\hbar\Omega_0\left\{\ketbra{\uparrow}{\downarrow}\otimes [a^\dagger\sin(\omega_bt)-a\sin(\omega_rt)]+h.c.\right\},
\end{eqnarray}
where the frequencies of the blue and red sideband drives satisfy the condition $\omega_b-\nu=\omega_r+\nu=\omega_0-\Delta$. By making a displacement transformation $a\to a+\eta$ and $a^\dagger \to a^\dagger+\eta$ ($\eta\ll 1$) and then moving to the interaction picture with respect to $H_0=\hbar(\omega_0-\Delta)/2\sigma_z+\hbar\nu a^\dagger a$, the effective Hamiltonian can be approximated as \cite{Supplement}
\begin{equation}
\label{eq:HI}
  H_I(t)\simeq\frac{\hbar\Delta}{2}\sigma_z-2i\hbar\eta\Omega_p(t)(a-a^\dagger)+\frac{\hbar\Omega_0}{2i}(a-a^\dagger)\sigma_x
\end{equation}
under the conditions of $\{|\Delta|,|\Omega_p(t)|,|\Omega_0|\}\ll\nu\ll\omega_0$. Therefore, it can be seen that the Hamiltonian in Eq.\eqref{eq:HI} as implemented in the trapped-ion platform is equivalent to the Hamiltonian in Eq.\eqref{eq:Haa} that governs the Dirac equation in the Alcubierre (1+1)-dimensional universe with the parameter correspondence as $v_s(t)\leftrightarrow 4\eta\Omega_p(t)/\Omega_0$, $c\leftrightarrow {(\hbar/2m_0 \nu)^{1/2}}\Omega_0/A$, and $m \leftrightarrow  -\hbar\Delta/2c^2$.
We remark that the experimental tunability of the parameters $\{\Omega_{0},\Omega_p,\Delta\}$ in the well-controllable trap-ion platform allows access to the crossover from flat to curved spacetimes, namely the tilt of the Dirac equation can be tuned by choosing appropriate values of the parameter $\Omega_{p}$, see Figure~\ref{fig:tilt_lightcones}. The angle of the simulated light cone is given by $\theta=\arctan\left(v_s+1\right)-\arctan\left(v_s-1\right)$, which decreases for a larger value of $v_s$, i.e. the velocity of the Dirac particle would be constrained in a smaller range. In the limit case of extremely large $v_s$, $\arctan\left(v_s\pm 1\right)\rightarrow \pi/2$ and the angle of the light cone $\theta\to 0$, the trajectory of the trapped ion can be seen as a counterpart of the closed timelike curves.
\section{Observation of analogue superluminal travel}
To show the effective velocity of the the simulated Dirac particle, we solve the following Heisenberg equation of the trapped ion system \cite{Supplement}
\begin{equation}\label{eq:velocity}
\begin{split}
\frac{d\hat{x}(t)}{dt}&=i\left[H(t),\hat{x}(t)\right]= c(v_s(t)+\hat{\phi}_z(t)),
\end{split}
\end{equation}
where $\hat{x}(t)=U^\dag(t)\hat{x}U(t)$ represents the position operator of the Dirac particle, and $\hat{\phi}_z(t)=U^\dag(t)\sigma_z U(t)$ with $U(t)$ the evolution operator from the Hamiltonian $H(t)$ in Eq.\eqref{eq:H}. We remark that Eq.\eqref{eq:velocity} is consistent with the Alcubierre metric, since the two eigenvalues $\pm1$ of $\hat{\phi}_z(t)$ correspond to the two opposite directions of the velocity. The mechanical degrees of freedom of the trapped ion $\hat{X}$ is related to the spatial coordinate of the simulated Dirac particle $\hat{x}$ by $\hat{X}\equiv A\hat{x}$ (see Eq.\ref{eq:XP}), thus the velocity $d \langle \hat{X}\rangle/dt$ is in the range of $Ac(v_s\pm 1)$.
\begin{figure}[t]
  \centering
  \includegraphics[width=6.5cm]{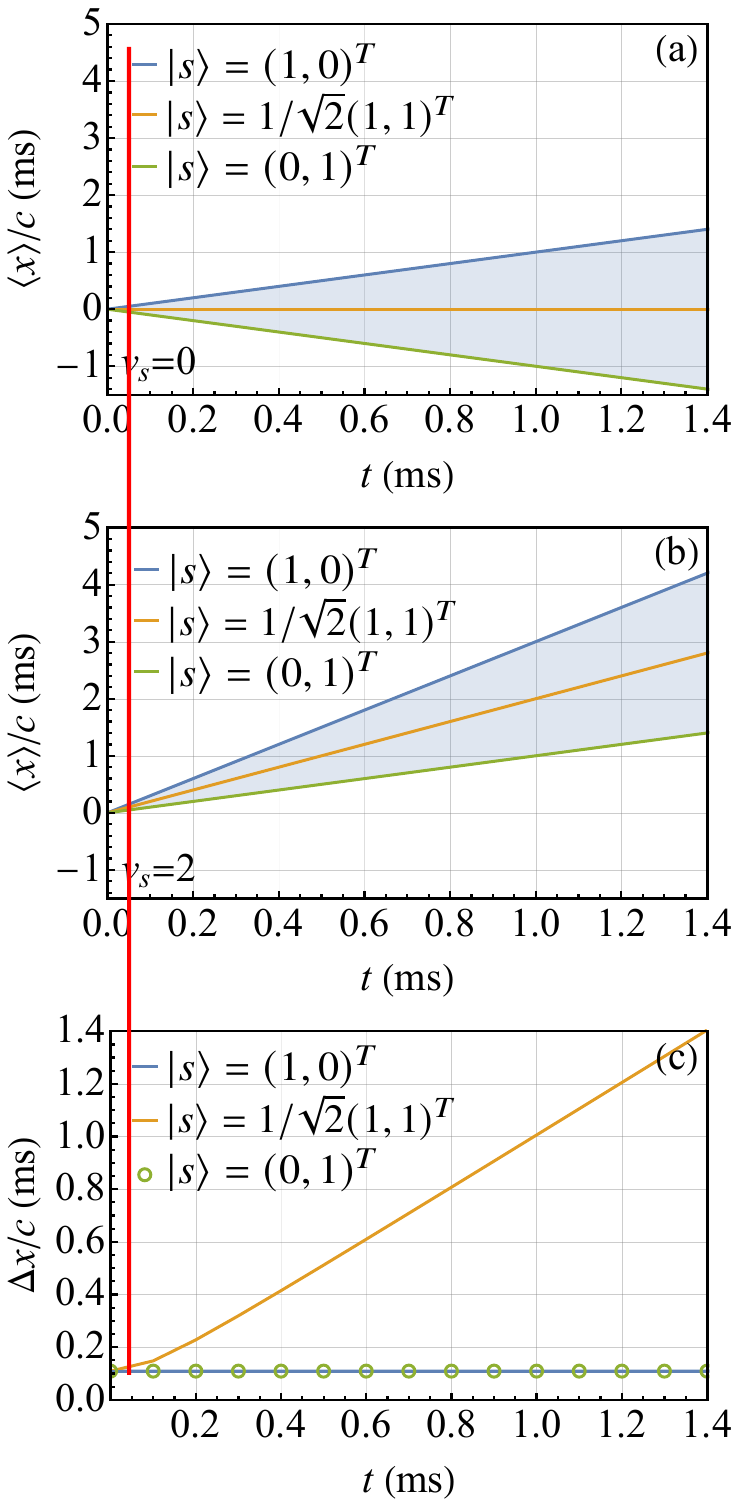}
  \caption{Dynamics of a trapped-ion simulated massless ($m=0$) Dirac particle in the Alcubierre spacetime. ({\bf a-b}) The light cones of the simulated Dirac particle (blue-shaded) are illustrated by its coordinate $\langle \hat{x}\rangle$, which is obtained from the average position of the trapped ion as $\langle \hat{x}\rangle=\langle \hat{X}\rangle/A$ for the Alcubierre metrics with ({\bf a}) $v_s=0$ (flat spacetime), and ({\bf b}) $v_s=2$ (curved spacetime). ({\bf c}) The variance of the simulated Dirac particle's position $\Delta x$ as a function of the evolution time for $v_s=2$. In ({\bf a-c}), the initial state of the trapped ion is chosen as $\ket{\psi_0}\sim\mathcal{N}\pi^{-1/4}\int dX\exp(-X^2/2)\ket{X}\otimes\ket{s}$, where $\ket{X}$ and $\ket{s}$ denote the spatial and internal degrees of freedom of the trapped ion. The internal states are chosen as $\ket{s}=(1,0)^T$ (blue), $\ket{s}=(0,1)^T$ (green), and $\ket{s}=1/\sqrt{2}(1,1)^T$ (orange). The parameters used in numerical calculations are $\Omega_0=2\pi\times1.46$ kHz, $\Omega_p=2\pi\times50$ kHz, $\nu=2\pi\times 5.9$ MHz, and $\Delta=0$. }
  \label{fig:m0}
\end{figure}
\begin{figure}[t]
  \centering
  \includegraphics[width=6.5cm]{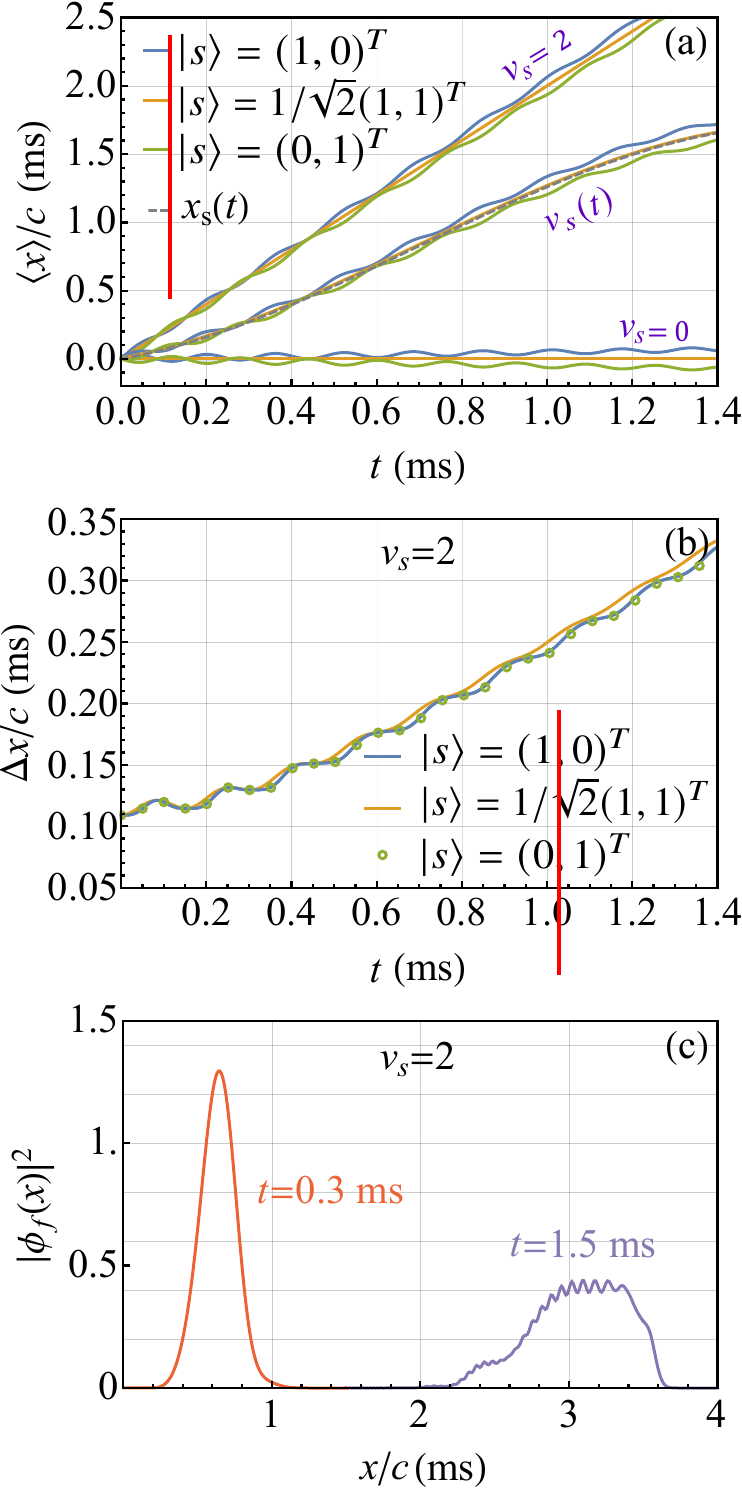}
  \caption{Observing Zitterbewegung effect in accompany with superluminal propagation for a trapped-ion simulated massive Dirac particle in the Alcubierre spacetime. ({\bf a}) The trajectories of the simulated Dirac particle with different initial internal states for $v_s=0$ (flat spacetime), $v_s=2$ (curved spacetime), and $v_s(t)=\partial_tx_s(t)$ where $x_s(t)=0.56t+1346t^2-642377t^3$. ({\bf b}) The variance of the simulated Dirac particle's position as a function of the evolution time $t$ for $v_s=2$. ({\bf c}) The wavepackets of the trapped ion at $t=0.3$ ms and $t=1.5$ ms as a function of the coordinate $x$ (in the unit of $c$). In ({\bf a-c}), the parameters used in numerical calculations are $\Omega_0=2\pi\times1.46$ kHz, $\nu=2\pi\times 5.9$ MHz, and $\Delta=-2\pi\times6.1$ kHz.}
  \label{fig:ZEffect}
\end{figure}
As an illustrative example, we choose the initial state of the trapped ion as $\ket{\psi_0}=\mathcal{N}\pi^{-1/4}\int dX\exp(-X^2/2)\ket{X}\otimes\ket{s}$, where $\ket{X}$ and $\ket{s}$ represent the spatial and internal degrees of freedom of the trapped ion respectively. Without loss of generality, the spatial wavefunction satisfies a Gaussian distribution with $\mathcal{N}$ the normalization factor. In Figure~\ref{fig:m0} (a) and (b), we show the trajectory $\langle \hat{x}(t)\rangle $ (in the unit of $c$) of the simulated massless Dirac particle (i.e. $m=0$)  as a function of the evolution time. According to Eq.~\eqref{eq:velocity}, the trajectory of the evolving states from the initial spin states $\ket{\uparrow}$ and $\ket{\downarrow}$ (namely the two eigenstates of $\langle \sigma_z\rangle =\pm 1$) specifies the shape of the simulated light cone. It can be seen that, corresponding to massless Dirac particles, the light cones for different values of the parameter $v_s$ are consistent with the analytical results from classical general relativity theory \cite{Garcia2019}. In particular, the simulated light cones start to tip over as the parameter $v_s$ increases, which indicates that the spacetime becomes more curved. These numerical simulation results demonstrate that the quantum simulation using trapped-ions can confirm the superluminal propagation of Dirac particles in the Alcubierre warp drive spacetime.
In the cases of initial states with $\ket{s}=\ket{\uparrow}$ or $\ket{\downarrow}$, the velocity is well defined, and thus the wavepacket of the simulated Dirac particle remains localized, as shown by the blue line and green circles in Figure~\ref{fig:m0} (c). In contrast, if we start from a coherent superposition state $\ket{s}=(\ket{0}+\ket{1})/\sqrt{2}$, the wavepacket of the simulated Dirac particle has two peaks
which correspond to the initial velocities $c(v_s\pm 1)$ respectively, resulting in an increase of the variance of the trajectory, see the orange line in Figure~\ref{fig:m0} (c).
\section{Zitterbewegung effect with superluminal propagation}
In addition to the analogue superluminal propagation, we also find that for massive Dirac particles with $m\ne 0$, the Zitterbewegung effect induces an oscillatory behaviour in the trajectory of the trapped ion. This effect is caused by the quantum superpositions of electron and positron solutions of the free wavepackets \cite{thaller2013,Lamata2007} and may exist in both flat and curved spacetime \cite{thaller2013,Pedernales2018}. The equation of motion for the expectation value of the position operator $\hat{x}(t)$ in this case can be obtained as \cite{Supplement}
\begin{equation}\label{eq:zeffect}
  \frac{d^2}{dt^2}\langle \hat{x}(t) \rangle =-\frac{2mc^3}{\hbar}\left\langle e^{2iH_\text{flat}t/\hbar}\sigma_y\right\rangle,
\end{equation}
where $H_\text{flat}=Ac\hat{P}\sigma_z-mc^2\sigma_x$ represents the Dirac Hamiltonian in the flat spacetime. From Eq.\eqref{eq:zeffect}, one can see that the oscillatory behaviour is induced by the mass $m$ but independent of the parameter $v_s$, and therefore the Zitterbewegung effect would persist with the superluminal propagation. In Figure~\ref{fig:ZEffect} (a), we show that the Zitterbewegung effect can manifest for both constant and time-dependent $v_s$ in the trapped-ion simulator. As an example, we consider a time-dependent spacetime by choosing $v_s$ such that both the initial and final velocities are subluminal, which is derived from a trajectory $x_s(t)$ of the simulated Dirac particle. This result is shown by the grey dashed line in Fig.\ref{fig:ZEffect} (a), which clearly exhibits the Zitterbewegung effect.
In Figure~\ref{fig:ZEffect} (b-c), we present the properties of the wavepacket of the simulated Dirac particle during its evolution. As shown in Figure~\ref{fig:ZEffect} (b), we do not observe any squeezing of the density profile caused by the simulated gravity in the trapped-ion system, which represents a different phenomenon from the one in Ref.\cite{Pedernales2018}. We remark that the qualitative behavior of the variance of the position is determined by the mass $m$ and is independent of the parameter $v_s$. The wavepackets at $t=0.3$ ms and $t=1.5$ ms are shown in Figure~\ref{fig:ZEffect}(c), which suggests that the wavepacket of the simulated Dirac particle is stretched and the oscillatory feature becomes more prominent as the system evolves, due to the Zitterbewegung effect.
{\it Conclusion.---} To summarize, we propose a scheme to implement a trapped-ion quantum simulation of a Dirac particle moving in the Alcubierre (1+1)-dimensional universe, which is a particularly important example of extreme curved spacetime leading to superluminal propagation in consistence with general relativity. We show that the flexibility of control parameters in such a platform allows one to observe counterintuitive effects including the superluminal velocity and the Zitterbewegung effect of the simulated Dirac particle in the Alcubierre curved spacetime. Our work demonstrates the feasibility to explore interesting and counterintuitive  features of exotic curved spacetime by quantum simulation of Dirac particles in the corresponding spacetime geometry. Furthermore, the analogy between particles in the setting of semiclassical quantum field theory and basic light-matter interaction quantum models would further inspire the ideas to investigate intriguing phenomena of general relativity and quantum field theory in a variety of quantum platforms that are available in laboratories.

{\it Acknowledgements.---} We thank Dr. Dongxiao Li and Prof. Jianwei Cui for very helpful discussions. This work is supported by National Natural Science Foundation of China (Grant No.~11690032), and the Open Project Program of Wuhan National Laboratory for Optoelectronics (No. 2019WNLOKF002).

\section{Appendix}

\subsection{Derivation of the Hamiltonian of a trapped-ion simulated Dirac particle}

In the main text, we propose to simulate a Dirac particle moving in the (1+1)-dimensional Alcubierre spacetime with a well-controlled trap-ion quantum simulator. In a semi-classical way, the Dirac particle moving in a fixed gravitational background, characterized by a general metric tensor $g_{\mu\nu}$, can be expressed as \cite{Mann1991,Collas2019}
\begin{equation}\label{eq:DE}
  \left(i\hbar\gamma^a e^\mu_{\,(a)}\partial_\mu+\frac{i\hbar}{2}\gamma^a\frac{1}{\sqrt{-g}}  \partial_\mu\left(\sqrt{-g}e^\mu_{\,(a)}\right)-mc^2\right)\psi=0,
\end{equation}
where the Greek letter $\mu$ and the Latin letter $a$ denote the coordinates of curved spacetime and local rest frame respectively, $m$ is the mass of the quantized field, $g$ is the determinant of the metric tensor, and $e^\mu_{\,(a)}$ is the vielbein allowing the constant Dirac matrices $\gamma^a$ to act at each spacetime point. Here, we choose the chiral representation such that $\gamma^0=\sigma_x$, $\gamma^1=-i\sigma_y$, with  ${\bm \sigma}$ the Pauli matrices. In the specific case of the Alcubierre spacetime, the metric tensor of which is given by

%i\hbar\partial_t\psi(t,x)=H\psi(t,x)
%with $t$ being the time of a distant observer.  %Hereafter, we set $\hbar\equiv1$ and $c\equiv1$.
\begin{equation}\label{eq:Ametric}
  g^{\mu\nu}=\left(
  \begin{matrix}
    1-v_s^2 & v_s \\
    v_s & -1
  \end{matrix}
  \right),
\end{equation}
the vielbein can be set as
\begin{equation}\label{eq:vielbein}
e^\mu_{\,(a)}=\left(
\begin{matrix}
  -1 & 0 \\
  -v_s(t) & -1
\end{matrix}
\right),
\end{equation}
with $a$ and $\mu$ the column  and row indices respectively. By multiplying Eq. \eqref{eq:DE} with $\sigma_z$, the Dirac equation can be expanded as
\begin{equation}\label{}
  \left[-mc^2\sigma_z+i\hbar\sigma_x\frac{\partial}{\partial x}-i\hbar\sigma_y\left(i v_s(t)\frac{\partial}{\partial x}+i\frac{\partial}{\partial (ct)}\right)\right]\psi(t,x)=0.
\end{equation}
which can be further rewritten in the form of Schr\"odinger equation [i.e. Eq. \textcolor{red}{(4)} in the main text] as follows
\begin{equation}\label{eq:S_EQ}
  i\hbar\partial_t\psi(t,x)=\left(-i c\hbar v_s(t)\frac{\partial}{\partial x}-i c\hbar\sigma_z\frac{\partial}{\partial x}-m c^2\sigma_x\right)\psi(t,x).
\end{equation}
Thus, the Hamiltonian of the Dirac particle to be simulated can be written as 
\begin{equation}
H(t)=cA\left(v_s(t)+\sigma_z\right)\hat{P}-m c^2\sigma_x,
\end{equation}
with $\hat{P}\equiv-i\hbar\partial/\partial \hat{X}$ and  $\hat{X}\equiv A\hat{x}$. We remark that the operators $\hat{X}$ and $\hat{P}$, which satisfy the standard commutation relation, i.e. $[\hat{X},\hat{P}]=i\hbar$, can be related to the position and momentum operators of the trapped ion used to simulate the relativistic Dirac particle. Thereby, we can rewrite them using the creation and annihilation operators as
\begin{equation}\label{eq:bo_field}
  \hat{X}=\sqrt{\frac{\hbar}{2m_0\nu}}(\hat{a}+\hat{a}^\dagger), \quad \hat{P}=-i\sqrt{\frac{m_0\nu}{2\hbar}}(\hat{a}-\hat{a}^\dagger),
\end{equation}
with $m_0$ the mass and $\nu$ the trapping potential of the trapped ion. In this way, the Hamiltonian of the trapped-ion simulated Dirac particle can be expressed as 
\begin{equation}\label{eq:Haa}
H(t)=\frac{c A \hbar}{i \sqrt{2}}\left(a-a^{\dagger}\right) \sigma_{z}+\frac{c A v_{s}(t) \hbar}{i \sqrt{2}}\left(a-a^{\dagger}\right)-m c^{2} \sigma_{x}.
\end{equation}
which is equivalent to the Eq. \textcolor{red}{(8)} in the main text by considering the Hadamard transformation.

\subsection{Quantum dynamics of the trapped-ion simulated Dirac particle}

In this section, we derive the quantum dynamics of the simulated Dirac particle in more detail. Based on the Heisenberg equation, the time evolution of the position and spin operators is given by
\begin{equation}\label{eq:velocity}
\begin{split}
   \frac{d\hat{x}(t)}{dt}&=\frac{i}{\hbar}\left[\tilde{H}(t),\hat{x}(t)\right] = c\left( v_s(t)+\hat{\phi}_z(t)\right),\\
   \frac{d\hat{\phi}_z(t)}{dt}& = -\frac{2mc^2}{\hbar}\hat{\phi}_y(t),
\end{split}
\end{equation}
where 
\begin{equation}
\hat{x}(t)=U^\dag(t)\hat{x}U(t),\quad \hat{\phi}_\alpha(t)=U^\dag(t)\sigma_\alpha U(t), 
\end{equation}
with $\alpha=x,y,z$.
The effective Hamiltonian and the evolution operator are 
\begin{equation}
\tilde{H}(t)=U^\dag(t)H(t) U(t), \quad U(t)=\exp[-i\int_0^t H(\tau)d\tau/\hbar],
\end{equation}
with $H(t)=cA\left(v_s(t)+\sigma_z\right)\hat{P}-m c^2\sigma_x$. In the above derivation, we have omitted the time ordering operator in $U(t)$ due to the fact that $[H(t),H(t^\prime)]=0$ for arbitrary $t$ and $t^\prime$. From the above equations, we can obtain
\begin{equation}
\begin{split}
\frac{d^2\hat{x}(t)}{dt^2}&=\frac{i}{\hbar}[H(t),c\hat{\phi}_z(t)]=-\frac{2mc^3}{\hbar}\hat{\phi}_y(t).
\end{split}
\end{equation}
In order to derive $\hat{\phi}_y(t)$, we notice that $\{\sigma_y,H(t)\}=2Acv_s(t)\hat{P}\sigma_y$, which gives $\sigma_yH(t)\sigma_y=-H(t)+2Acv_s(t)\hat{P}$ and further
\begin{equation}
\begin{split}
  \hat{\phi}_y(t)=&\exp\left(i\int H(t)dt/\hbar\right)\sigma_y\exp\left(-i\int H(t)dt/\hbar\right)\\
  =&\exp\left(i\int H(t)dt/\hbar\right)\exp\left(-i\int \sigma_yH(t)\sigma_ydt/\hbar\right)\sigma_y\\
  =&\exp\left[2i\int\left(H(t)-Acv_s(t)\hat{P}\right)dt\right]\sigma_y\\
  \equiv & \exp\left(2iH_\text{flat}t/\hbar\right)\sigma_y,
\end{split}
\end{equation}
where $H_\text{flat}=Ac\hat{P}\sigma_z-mc^2\sigma_x$ corresponds to the Dirac Hamiltonian in the flat spacetime. Hence,
we have [i.e. Eq. \textcolor{red}{(12)} in the main text]
\begin{equation}\label{eq:acceleration}
\begin{split}
\frac{d^2\left\langle\hat{x}(t)\right\rangle}{dt^2}= -\frac{2mc^3}{\hbar}\left\langle e^{2iH_\text{flat}t/\hbar}\sigma_y\right\rangle.
\end{split}
\end{equation}

\subsection{Trapped-ion realization of the Dirac Hamiltonian}
The well-controllable trapped-ion system is a particularly promising platform for simulating the basic light-matter interaction models by exploiting its internal and motional degrees of freedom. Here we explain in more detail the trapped-ion setup to simulate the Dirac particle moving in a curved spacetime, the Hamiltonian of which (i.e. Eq. \textcolor{red}{(8)} in the main text) has been derived in the above section. We consider the radial motional mode with frequency $\nu$ of the trapped ion to model the bosonic field and the two internal ground hyperfine levels with transition frequency $\omega_0$ to form an effective spin-1/2, denoted as $|{\uparrow}\rangle$ and $|{\downarrow}\rangle$. With blue and red sideband drives at frequencies $\omega_b$ and $\omega_r$ respectively \cite{Sorensen1999,Burd2021}, which can be generated by using oscillating near-field magnetic field gradients in experiments  \cite{Srinivas2019}, one is able to realize the spin-harmonic oscillator coupling via the Mølmer–Sørensen interaction that is described as 
\begin{equation}
H_{\Omega_0}=\hbar \Omega_{0}\left\{\left|\uparrow\right\rangle\left\langle\downarrow\right| \otimes\left[a^{\dagger} \sin \left(\omega_{b} t\right)-a \sin \left(\omega_{r} t\right)\right]+h . c .\right\}.
\end{equation}
In addition, it is possible to apply an oscillating potential directly to the radio-frequency trapping electrodes such that the single ion’s trapping potential is modulated periodically with an amplitude  $\Omega_p(t)$ at frequency $\nu$. The corresponding Hamitonian is given by
\begin{equation}
H_{\Omega_p}=\hbar \Omega_{p}(t) \sin (v t)\left(a+a^{\dagger}\right)^{2}
\end{equation}
Therefore, the total Hamiltonian of such a trapped-ion setup is [namely Eq. \textcolor{red}{(9)} in the main text]
\begin{equation}
  \label{eq:H_lab}
  \begin{split}
H_\text{ion}&= \frac{\hbar \omega_{0}}{2} \sigma_{z}+\hbar v a^{\dagger} a+H_{\Omega_p}+H_{\Omega_0}\\
            &=\frac{\hbar \omega_{0}}{2} \sigma_{z}+\hbar v a^{\dagger} a+\hbar \Omega_{p}(t) \sin (\nu t)\left(a+a^{\dagger}\right)^{2}\\
  &+\hbar \Omega_{0}\left\{\left|\uparrow\right\rangle\left\langle\downarrow\right| \otimes\left[a^{\dagger} \sin \left(\omega_{b} t\right)-a \sin \left(\omega_{r} t\right)\right]+h . c .\right\}.
\end{split}
\end{equation}
In order to derive the Hamiltonian of Eq. \textcolor{red}{(8)} in the main text, we first make a displace transformation to the above Hamiltonian with $\mathcal{D}[\eta]=e^{\eta(a^\dagger-a)}$, $\eta\ll 1$, which gives
\begin{equation}\label{}
\begin{split}
  H_\text{ion}'&=\mathcal{D}^\dagger[\eta]H_\text{ion}\mathcal{D}[\eta]\\
  &=\,\frac{\hbar\omega_0}{2}\sigma_z+\hbar\nu a^\dagger a+\hbar\nu\eta(a+a^\dagger)\\
  &+\hbar\Omega_0\left\{\ketbra{\uparrow}{\downarrow}[(a^\dagger+\eta) \sin(\omega_bt) -(a+\eta)\sin(\omega_rt)]+h.c.\right\}\\
  &+\hbar\Omega_p\sin (\nu t)[a^2+a^{\dagger2}+aa^\dagger+a^\dagger a+4\eta(a+a^\dagger)+4\eta^2].
\end{split}
\end{equation}
Then we move to the interaction picture with respect to $H_0=\hbar(\omega_0-\Delta)/2\sigma_z+\hbar\nu a^\dagger a$ while requiring $\omega_b-\nu=\omega_r+\nu=\omega_0-\Delta$. After applying the rotating-wave approximation under the conditions of $\{|\Delta|,|\Omega_p(t)|,|\Omega_0|\}\ll\nu\ll\omega_0$, we are left with
\begin{equation}\label{eq:HI}
  H_{I,\text{ion}}\simeq \frac{\hbar\Delta}{2}\sigma_z +\frac{\hbar\Omega_0}{2i}\left[\ketbra{\uparrow}{\downarrow}(a-a^\dagger) +h.c.\right] -2i\eta\hbar\Omega_p(a-a^\dagger),
\end{equation}
the form of which is equivalent to the Hamiltonian of Eq. \textcolor{red}{(10)} in the main text by further making the Hadamard transformation. 
\subsection{Numerical simulation of the trapped-ion simulated Dirac particle}

Without loss of generality, the initial state of the trapped ion is chosen as 
\begin{equation}
\ket{\psi_0}=\mathcal{N}\pi^{-1/4}\int dX\exp(-X^2/2)\ket{X}\ket{s}, 
\end{equation}
where $\ket{X}$ and $\ket{s}$ denote the spatial and internal degrees of freedom of the trapped ion respectively, and $\mathcal{N}$ is the normalization factor of its wavefunction. Using the following relation 
\begin{equation}
 \braket{X}{n}=(\sqrt{2}/2)^n(n!)^{-1/2}\pi^{-1/4}e^{-X^2/2}\text{H}_n(X),
\end{equation}
where $\text{H}_n(X)$ are the Hermite polynomials, we can rewrite the initial state in the Fock space as
\begin{equation}
\ket{\psi_0}=\sum_nc_n\ket{n}\ket{s} 
\end{equation}
with $c_n=\mathcal{N}\pi^{-1/4}\int dX\exp(-X^2/2)\braket{n}{X}$. Hence, the average position 
\begin{equation}
\langle x\rangle=\bra{\psi_0}U^\dagger\hat{x}U\ket{\psi_0},
\end{equation}
and the corresponding variance 
\begin{equation}
\Delta x=\left(\bra{\psi_0}U^\dagger\hat{x}^2U\ket{\psi_0}-\bra{\psi_0}U^\dagger\hat{x}U\ket{\psi_0}^2\right)^{1/2}
\end{equation}
can be obtained with respect to the time evolved state $U(t)\ket{\psi_0}$ in a truncated Fock space with the maximum excitation number $n_{\text{max}}=512$. We remark that this truncation is verified to guarantee converging results, which are shown in Fig.\textcolor{red}{2} and Fig.\textcolor{red}{3 (a-b)} in the main text. In addition, the probability distribution of the wavepacket $|\phi_f(x)|^2$ corresponding to the final state can be obtained by
\begin{equation}\label{}
\begin{split}
  \left|\phi_f(x)\right|^2&=\text{Tr}\left(\rho_x\ketbra{x}{x}\right)=\text{Tr}\left(\rho_x\sum_{n,n'}\ket{n}\braket{n}{x}\braket{x}{n'}\bra{n'}\right),
\end{split}
\end{equation}
where $\rho_x$ is the reduced density matrix for the spatial degrees of freedom by
tracing out the spin degree of freedom in the total final density matrix of the final state $U(t)\ket{\psi_0}\bra{\psi_0}U^\dagger(t)$.

\bibliography{literature}

\begin{thebibliography}{56}%
\makeatletter
\providecommand \@ifxundefined [1]{%
 \@ifx{#1\undefined}
}%
\providecommand \@ifnum [1]{%
 \ifnum #1\expandafter \@firstoftwo
 \else \expandafter \@secondoftwo
 \fi
}%
\providecommand \@ifx [1]{%
 \ifx #1\expandafter \@firstoftwo
 \else \expandafter \@secondoftwo
 \fi
}%
\providecommand \natexlab [1]{#1}%
\providecommand \enquote  [1]{``#1''}%
\providecommand \bibnamefont  [1]{#1}%
\providecommand \bibfnamefont [1]{#1}%
\providecommand \citenamefont [1]{#1}%
\providecommand \href@noop [0]{\@secondoftwo}%
\providecommand \href [0]{\begingroup \@sanitize@url \@href}%
\providecommand \@href[1]{\@@startlink{#1}\@@href}%
\providecommand \@@href[1]{\endgroup#1\@@endlink}%
\providecommand \@sanitize@url [0]{\catcode `\\12\catcode `\$12\catcode
  `\&12\catcode `\#12\catcode `\^12\catcode `\_12\catcode `\%12\relax}%
\providecommand \@@startlink[1]{}%
\providecommand \@@endlink[0]{}%
\providecommand \url  [0]{\begingroup\@sanitize@url \@url }%
\providecommand \@url [1]{\endgroup\@href {#1}{\urlprefix }}%
\providecommand \urlprefix  [0]{URL }%
\providecommand \Eprint [0]{\href }%
\providecommand \doibase [0]{https://doi.org/}%
\providecommand \selectlanguage [0]{\@gobble}%
\providecommand \bibinfo  [0]{\@secondoftwo}%
\providecommand \bibfield  [0]{\@secondoftwo}%
\providecommand \translation [1]{[#1]}%
\providecommand \BibitemOpen [0]{}%
\providecommand \bibitemStop [0]{}%
\providecommand \bibitemNoStop [0]{.\EOS\space}%
\providecommand \EOS [0]{\spacefactor3000\relax}%
\providecommand \BibitemShut  [1]{\csname bibitem#1\endcsname}%
\let\auto@bib@innerbib\@empty
%</preamble>
\bibitem [{\citenamefont {Davies}(1976)}]{Davis1976}%
  \BibitemOpen
  \bibfield  {author} {\bibinfo {author} {\bibfnamefont {P.~C.~W.}\
  \bibnamefont {Davies}},\ }\bibfield  {title} {\bibinfo {title} {Quantum field
  theory in curved space–time},\ }\href {https://doi.org/10.1038/263377a0}
  {\bibfield  {journal} {\bibinfo  {journal} {Nature}\ }\textbf {\bibinfo
  {volume} {263}},\ \bibinfo {pages} {377} (\bibinfo {year}
  {1976})}\BibitemShut {NoStop}%
\bibitem [{\citenamefont {Jacobson}(2005)}]{Jacobson2005}%
  \BibitemOpen
  \bibfield  {author} {\bibinfo {author} {\bibfnamefont {T.}~\bibnamefont
  {Jacobson}},\ }\bibfield  {title} {\bibinfo {title} {Introduction to quantum
  fields in curved spacetime and the {H}awking effect},\ }in\ \href@noop {}
  {\emph {\bibinfo {booktitle} {Lectures on Quantum Gravity}}}\ (\bibinfo
  {publisher} {Springer},\ \bibinfo {year} {2005})\ pp.\ \bibinfo {pages}
  {39--89}\BibitemShut {NoStop}%
\bibitem [{\citenamefont {Hollands}\ and\ \citenamefont
  {Wald}(2015)}]{Hollands2015}%
  \BibitemOpen
  \bibfield  {author} {\bibinfo {author} {\bibfnamefont {S.}~\bibnamefont
  {Hollands}}\ and\ \bibinfo {author} {\bibfnamefont {R.~M.}\ \bibnamefont
  {Wald}},\ }\bibfield  {title} {\bibinfo {title} {Quantum fields in curved
  spacetime},\ }\href
  {https://doi.org/https://doi.org/10.1016/j.physrep.2015.02.001} {\bibfield
  {journal} {\bibinfo  {journal} {Physics Reports}\ }\textbf {\bibinfo {volume}
  {574}},\ \bibinfo {pages} {1 } (\bibinfo {year} {2015})}\BibitemShut
  {NoStop}%
\bibitem [{\citenamefont {Howl}\ \emph {et~al.}(2018)\citenamefont {Howl},
  \citenamefont {Hackermülller}, \citenamefont {Bruschi},\ and\ \citenamefont
  {Fuentes}}]{Howl2018}%
  \BibitemOpen
  \bibfield  {author} {\bibinfo {author} {\bibfnamefont {R.}~\bibnamefont
  {Howl}}, \bibinfo {author} {\bibfnamefont {L.}~\bibnamefont
  {Hackermülller}}, \bibinfo {author} {\bibfnamefont {D.~E.}\ \bibnamefont
  {Bruschi}},\ and\ \bibinfo {author} {\bibfnamefont {I.}~\bibnamefont
  {Fuentes}},\ }\bibfield  {title} {\bibinfo {title} {Gravity in the quantum
  lab},\ }\href {https://doi.org/10.1080/23746149.2017.1383184} {\bibfield
  {journal} {\bibinfo  {journal} {Adv. Phys.: X}\ }\textbf {\bibinfo {volume}
  {3}},\ \bibinfo {pages} {1383184} (\bibinfo {year} {2018})}\BibitemShut
  {NoStop}%
\bibitem [{\citenamefont {Howl}\ \emph {et~al.}(2021)\citenamefont {Howl},
  \citenamefont {Vedral}, \citenamefont {Naik}, \citenamefont {Christodoulou},
  \citenamefont {Rovelli},\ and\ \citenamefont {Iyer}}]{Howl2021}%
  \BibitemOpen
  \bibfield  {author} {\bibinfo {author} {\bibfnamefont {R.}~\bibnamefont
  {Howl}}, \bibinfo {author} {\bibfnamefont {V.}~\bibnamefont {Vedral}},
  \bibinfo {author} {\bibfnamefont {D.}~\bibnamefont {Naik}}, \bibinfo {author}
  {\bibfnamefont {M.}~\bibnamefont {Christodoulou}}, \bibinfo {author}
  {\bibfnamefont {C.}~\bibnamefont {Rovelli}},\ and\ \bibinfo {author}
  {\bibfnamefont {A.}~\bibnamefont {Iyer}},\ }\bibfield  {title} {\bibinfo
  {title} {Non-gaussianity as a signature of a quantum theory of gravity},\
  }\href {https://doi.org/10.1103/PRXQuantum.2.010325} {\bibfield  {journal}
  {\bibinfo  {journal} {PRX Quantum}\ }\textbf {\bibinfo {volume} {2}},\
  \bibinfo {pages} {010325} (\bibinfo {year} {2021})}\BibitemShut {NoStop}%
\bibitem [{\citenamefont {Unruh}(1981)}]{Unruh1981}%
  \BibitemOpen
  \bibfield  {author} {\bibinfo {author} {\bibfnamefont {W.~G.}\ \bibnamefont
  {Unruh}},\ }\bibfield  {title} {\bibinfo {title} {Experimental black-hole
  evaporation?},\ }\href {https://doi.org/10.1103/PhysRevLett.46.1351}
  {\bibfield  {journal} {\bibinfo  {journal} {Phys. Rev. Lett.}\ }\textbf
  {\bibinfo {volume} {46}},\ \bibinfo {pages} {1351} (\bibinfo {year}
  {1981})}\BibitemShut {NoStop}%
\bibitem [{\citenamefont {Weinfurtner}\ \emph {et~al.}(2011)\citenamefont
  {Weinfurtner}, \citenamefont {Tedford}, \citenamefont {Penrice},
  \citenamefont {Unruh},\ and\ \citenamefont {Lawrence}}]{Weinfurtner2011}%
  \BibitemOpen
  \bibfield  {author} {\bibinfo {author} {\bibfnamefont {S.}~\bibnamefont
  {Weinfurtner}}, \bibinfo {author} {\bibfnamefont {E.~W.}\ \bibnamefont
  {Tedford}}, \bibinfo {author} {\bibfnamefont {M.~C.~J.}\ \bibnamefont
  {Penrice}}, \bibinfo {author} {\bibfnamefont {W.~G.}\ \bibnamefont {Unruh}},\
  and\ \bibinfo {author} {\bibfnamefont {G.~A.}\ \bibnamefont {Lawrence}},\
  }\bibfield  {title} {\bibinfo {title} {Measurement of stimulated hawking
  emission in an analogue system},\ }\href
  {https://doi.org/10.1103/PhysRevLett.106.021302} {\bibfield  {journal}
  {\bibinfo  {journal} {Phys. Rev. Lett.}\ }\textbf {\bibinfo {volume} {106}},\
  \bibinfo {pages} {021302} (\bibinfo {year} {2011})}\BibitemShut {NoStop}%
\bibitem [{\citenamefont {Zurek}(1985)}]{Zurek1985}%
  \BibitemOpen
  \bibfield  {author} {\bibinfo {author} {\bibfnamefont {W.~H.}\ \bibnamefont
  {Zurek}},\ }\bibfield  {title} {\bibinfo {title} {Cosmological experiments in
  superfluid helium?},\ }\href@noop {} {\bibfield  {journal} {\bibinfo
  {journal} {Nature}\ }\textbf {\bibinfo {volume} {317}},\ \bibinfo {pages}
  {505} (\bibinfo {year} {1985})}\BibitemShut {NoStop}%
\bibitem [{\citenamefont {Fedichev}\ and\ \citenamefont
  {Fischer}(2004)}]{Fedichev2004}%
  \BibitemOpen
  \bibfield  {author} {\bibinfo {author} {\bibfnamefont {P.~O.}\ \bibnamefont
  {Fedichev}}\ and\ \bibinfo {author} {\bibfnamefont {U.~R.}\ \bibnamefont
  {Fischer}},\ }\bibfield  {title} {\bibinfo {title} {``cosmological''
  quasiparticle production in harmonically trapped superfluid gases},\ }\href
  {https://doi.org/10.1103/PhysRevA.69.033602} {\bibfield  {journal} {\bibinfo
  {journal} {Phys. Rev. A}\ }\textbf {\bibinfo {volume} {69}},\ \bibinfo
  {pages} {033602} (\bibinfo {year} {2004})}\BibitemShut {NoStop}%
\bibitem [{\citenamefont {Garay}\ \emph {et~al.}(2000)\citenamefont {Garay},
  \citenamefont {Anglin}, \citenamefont {Cirac},\ and\ \citenamefont
  {Zoller}}]{Garay2000}%
  \BibitemOpen
  \bibfield  {author} {\bibinfo {author} {\bibfnamefont {L.~J.}\ \bibnamefont
  {Garay}}, \bibinfo {author} {\bibfnamefont {J.~R.}\ \bibnamefont {Anglin}},
  \bibinfo {author} {\bibfnamefont {J.~I.}\ \bibnamefont {Cirac}},\ and\
  \bibinfo {author} {\bibfnamefont {P.}~\bibnamefont {Zoller}},\ }\bibfield
  {title} {\bibinfo {title} {Sonic analog of gravitational black holes in
  {B}ose-{E}instein condensates},\ }\href
  {https://doi.org/10.1103/PhysRevLett.85.4643} {\bibfield  {journal} {\bibinfo
   {journal} {Phys. Rev. Lett.}\ }\textbf {\bibinfo {volume} {85}},\ \bibinfo
  {pages} {4643} (\bibinfo {year} {2000})}\BibitemShut {NoStop}%
\bibitem [{\citenamefont {Barcel\'o}\ \emph {et~al.}(2003)\citenamefont
  {Barcel\'o}, \citenamefont {Liberati},\ and\ \citenamefont
  {Visser}}]{Barcelo2003}%
  \BibitemOpen
  \bibfield  {author} {\bibinfo {author} {\bibfnamefont {C.}~\bibnamefont
  {Barcel\'o}}, \bibinfo {author} {\bibfnamefont {S.}~\bibnamefont
  {Liberati}},\ and\ \bibinfo {author} {\bibfnamefont {M.}~\bibnamefont
  {Visser}},\ }\bibfield  {title} {\bibinfo {title} {Probing semiclassical
  analog gravity in {B}ose-{E}instein condensates with widely tunable
  interactions},\ }\href {https://doi.org/10.1103/PhysRevA.68.053613}
  {\bibfield  {journal} {\bibinfo  {journal} {Phys. Rev. A}\ }\textbf {\bibinfo
  {volume} {68}},\ \bibinfo {pages} {053613} (\bibinfo {year}
  {2003})}\BibitemShut {NoStop}%
\bibitem [{\citenamefont {Fedichev}\ and\ \citenamefont
  {Fischer}(2003)}]{Fedichev2003}%
  \BibitemOpen
  \bibfield  {author} {\bibinfo {author} {\bibfnamefont {P.~O.}\ \bibnamefont
  {Fedichev}}\ and\ \bibinfo {author} {\bibfnamefont {U.~R.}\ \bibnamefont
  {Fischer}},\ }\bibfield  {title} {\bibinfo {title} {Gibbons-{H}awking effect
  in the sonic de sitter space-time of an expanding {B}ose-{E}instein-condensed
  gas},\ }\href {https://doi.org/10.1103/PhysRevLett.91.240407} {\bibfield
  {journal} {\bibinfo  {journal} {Phys. Rev. Lett.}\ }\textbf {\bibinfo
  {volume} {91}},\ \bibinfo {pages} {240407} (\bibinfo {year}
  {2003})}\BibitemShut {NoStop}%
\bibitem [{\citenamefont {Fischer}\ and\ \citenamefont
  {Sch\"utzhold}(2004)}]{Fischer2004}%
  \BibitemOpen
  \bibfield  {author} {\bibinfo {author} {\bibfnamefont {U.~R.}\ \bibnamefont
  {Fischer}}\ and\ \bibinfo {author} {\bibfnamefont {R.}~\bibnamefont
  {Sch\"utzhold}},\ }\bibfield  {title} {\bibinfo {title} {Quantum simulation
  of cosmic inflation in two-component {B}ose-{E}instein condensates},\ }\href
  {https://doi.org/10.1103/PhysRevA.70.063615} {\bibfield  {journal} {\bibinfo
  {journal} {Phys. Rev. A}\ }\textbf {\bibinfo {volume} {70}},\ \bibinfo
  {pages} {063615} (\bibinfo {year} {2004})}\BibitemShut {NoStop}%
\bibitem [{\citenamefont {Carusotto}\ \emph {et~al.}(2008)\citenamefont
  {Carusotto}, \citenamefont {Fagnocchi}, \citenamefont {Recati}, \citenamefont
  {Balbinot},\ and\ \citenamefont {Fabbri}}]{Carusotto2008}%
  \BibitemOpen
  \bibfield  {author} {\bibinfo {author} {\bibfnamefont {I.}~\bibnamefont
  {Carusotto}}, \bibinfo {author} {\bibfnamefont {S.}~\bibnamefont
  {Fagnocchi}}, \bibinfo {author} {\bibfnamefont {A.}~\bibnamefont {Recati}},
  \bibinfo {author} {\bibfnamefont {R.}~\bibnamefont {Balbinot}},\ and\
  \bibinfo {author} {\bibfnamefont {A.}~\bibnamefont {Fabbri}},\ }\bibfield
  {title} {\bibinfo {title} {Numerical observation of {H}awking radiation from
  acoustic black holes in atomic {B}ose{\textendash}{E}instein condensates},\
  }\href {https://doi.org/10.1088/1367-2630/10/10/103001} {\bibfield  {journal}
  {\bibinfo  {journal} {New Journal of Physics}\ }\textbf {\bibinfo {volume}
  {10}},\ \bibinfo {pages} {103001} (\bibinfo {year} {2008})}\BibitemShut
  {NoStop}%
\bibitem [{\citenamefont {Steinhauer}(2014)}]{steinhauer2014observation}%
  \BibitemOpen
  \bibfield  {author} {\bibinfo {author} {\bibfnamefont {J.}~\bibnamefont
  {Steinhauer}},\ }\bibfield  {title} {\bibinfo {title} {Observation of
  self-amplifying {H}awking radiation in an analogue black-hole laser},\ }\href
  {https://www.nature.com/articles/nphys3104} {\bibfield  {journal} {\bibinfo
  {journal} {Nature Physics}\ }\textbf {\bibinfo {volume} {10}},\ \bibinfo
  {pages} {864} (\bibinfo {year} {2014})}\BibitemShut {NoStop}%
\bibitem [{\citenamefont {Steinhauer}(2016)}]{Steinhauer2016}%
  \BibitemOpen
  \bibfield  {author} {\bibinfo {author} {\bibfnamefont {J.}~\bibnamefont
  {Steinhauer}},\ }\bibfield  {title} {\bibinfo {title} {Observation of quantum
  {H}awking radiation and its entanglement in an analogue black hole},\ }\href
  {https://www.nature.com/articles/nphys3863} {\bibfield  {journal} {\bibinfo
  {journal} {Nature Physics}\ }\textbf {\bibinfo {volume} {12}},\ \bibinfo
  {pages} {959} (\bibinfo {year} {2016})}\BibitemShut {NoStop}%
\bibitem [{\citenamefont {Ch\"a}\ and\ \citenamefont
  {Fischer}(2017)}]{Cha2017}%
  \BibitemOpen
  \bibfield  {author} {\bibinfo {author} {\bibfnamefont {S.-Y.}\ \bibnamefont
  {Ch\"a}}\ and\ \bibinfo {author} {\bibfnamefont {U.~R.}\ \bibnamefont
  {Fischer}},\ }\bibfield  {title} {\bibinfo {title} {Probing the scale
  invariance of the inflationary power spectrum in expanding
  quasi-two-dimensional dipolar condensates},\ }\href
  {https://doi.org/10.1103/PhysRevLett.118.130404} {\bibfield  {journal}
  {\bibinfo  {journal} {Phys. Rev. Lett.}\ }\textbf {\bibinfo {volume} {118}},\
  \bibinfo {pages} {130404} (\bibinfo {year} {2017})}\BibitemShut {NoStop}%
\bibitem [{\citenamefont {De~Nova}\ \emph {et~al.}(2019)\citenamefont
  {De~Nova}, \citenamefont {Golubkov}, \citenamefont {Kolobov},\ and\
  \citenamefont {Steinhauer}}]{De2019}%
  \BibitemOpen
  \bibfield  {author} {\bibinfo {author} {\bibfnamefont {J.~R.~M.}\
  \bibnamefont {De~Nova}}, \bibinfo {author} {\bibfnamefont {K.}~\bibnamefont
  {Golubkov}}, \bibinfo {author} {\bibfnamefont {V.~I.}\ \bibnamefont
  {Kolobov}},\ and\ \bibinfo {author} {\bibfnamefont {J.}~\bibnamefont
  {Steinhauer}},\ }\bibfield  {title} {\bibinfo {title} {Observation of thermal
  {H}awking radiation and its temperature in an analogue black hole},\ }\href
  {https://www.nature.com/articles/s41586-019-1241-0} {\bibfield  {journal}
  {\bibinfo  {journal} {Nature}\ }\textbf {\bibinfo {volume} {569}},\ \bibinfo
  {pages} {688} (\bibinfo {year} {2019})}\BibitemShut {NoStop}%
\bibitem [{\citenamefont {Isoard}\ and\ \citenamefont
  {Pavloff}(2020)}]{Isoard2020}%
  \BibitemOpen
  \bibfield  {author} {\bibinfo {author} {\bibfnamefont {M.}~\bibnamefont
  {Isoard}}\ and\ \bibinfo {author} {\bibfnamefont {N.}~\bibnamefont
  {Pavloff}},\ }\bibfield  {title} {\bibinfo {title} {Departing from thermality
  of analogue {H}awking radiation in a {B}ose-{E}instein condensate},\ }\href
  {https://doi.org/10.1103/PhysRevLett.124.060401} {\bibfield  {journal}
  {\bibinfo  {journal} {Phys. Rev. Lett.}\ }\textbf {\bibinfo {volume} {124}},\
  \bibinfo {pages} {060401} (\bibinfo {year} {2020})}\BibitemShut {NoStop}%
\bibitem [{\citenamefont {Belgiorno}\ \emph {et~al.}(2010)\citenamefont
  {Belgiorno}, \citenamefont {Cacciatori}, \citenamefont {Clerici},
  \citenamefont {Gorini}, \citenamefont {Ortenzi}, \citenamefont {Rizzi},
  \citenamefont {Rubino}, \citenamefont {Sala},\ and\ \citenamefont
  {Faccio}}]{Belgiorno2010}%
  \BibitemOpen
  \bibfield  {author} {\bibinfo {author} {\bibfnamefont {F.}~\bibnamefont
  {Belgiorno}}, \bibinfo {author} {\bibfnamefont {S.~L.}\ \bibnamefont
  {Cacciatori}}, \bibinfo {author} {\bibfnamefont {M.}~\bibnamefont {Clerici}},
  \bibinfo {author} {\bibfnamefont {V.}~\bibnamefont {Gorini}}, \bibinfo
  {author} {\bibfnamefont {G.}~\bibnamefont {Ortenzi}}, \bibinfo {author}
  {\bibfnamefont {L.}~\bibnamefont {Rizzi}}, \bibinfo {author} {\bibfnamefont
  {E.}~\bibnamefont {Rubino}}, \bibinfo {author} {\bibfnamefont {V.~G.}\
  \bibnamefont {Sala}},\ and\ \bibinfo {author} {\bibfnamefont
  {D.}~\bibnamefont {Faccio}},\ }\bibfield  {title} {\bibinfo {title}
  {{H}awking radiation from ultrashort laser pulse filaments},\ }\href
  {https://doi.org/10.1103/PhysRevLett.105.203901} {\bibfield  {journal}
  {\bibinfo  {journal} {Phys. Rev. Lett.}\ }\textbf {\bibinfo {volume} {105}},\
  \bibinfo {pages} {203901} (\bibinfo {year} {2010})}\BibitemShut {NoStop}%
\bibitem [{\citenamefont {Philbin}\ \emph {et~al.}(2008)\citenamefont
  {Philbin}, \citenamefont {Kuklewicz}, \citenamefont {Robertson},
  \citenamefont {Hill}, \citenamefont {K{\"o}nig},\ and\ \citenamefont
  {Leonhardt}}]{Philbin2008}%
  \BibitemOpen
  \bibfield  {author} {\bibinfo {author} {\bibfnamefont {T.~G.}\ \bibnamefont
  {Philbin}}, \bibinfo {author} {\bibfnamefont {C.}~\bibnamefont {Kuklewicz}},
  \bibinfo {author} {\bibfnamefont {S.}~\bibnamefont {Robertson}}, \bibinfo
  {author} {\bibfnamefont {S.}~\bibnamefont {Hill}}, \bibinfo {author}
  {\bibfnamefont {F.}~\bibnamefont {K{\"o}nig}},\ and\ \bibinfo {author}
  {\bibfnamefont {U.}~\bibnamefont {Leonhardt}},\ }\bibfield  {title} {\bibinfo
  {title} {Fiber-optical analog of the event horizon},\ }\href
  {https://doi.org/10.1126/science.1153625} {\bibfield  {journal} {\bibinfo
  {journal} {Science}\ }\textbf {\bibinfo {volume} {319}},\ \bibinfo {pages}
  {1367} (\bibinfo {year} {2008})}\BibitemShut {NoStop}%
\bibitem [{\citenamefont {Drori}\ \emph {et~al.}(2019)\citenamefont {Drori},
  \citenamefont {Rosenberg}, \citenamefont {Bermudez}, \citenamefont
  {Silberberg},\ and\ \citenamefont {Leonhardt}}]{Drori2019}%
  \BibitemOpen
  \bibfield  {author} {\bibinfo {author} {\bibfnamefont {J.}~\bibnamefont
  {Drori}}, \bibinfo {author} {\bibfnamefont {Y.}~\bibnamefont {Rosenberg}},
  \bibinfo {author} {\bibfnamefont {D.}~\bibnamefont {Bermudez}}, \bibinfo
  {author} {\bibfnamefont {Y.}~\bibnamefont {Silberberg}},\ and\ \bibinfo
  {author} {\bibfnamefont {U.}~\bibnamefont {Leonhardt}},\ }\bibfield  {title}
  {\bibinfo {title} {Observation of stimulated {H}awking radiation in an
  optical analogue},\ }\href {https://doi.org/10.1103/PhysRevLett.122.010404}
  {\bibfield  {journal} {\bibinfo  {journal} {Phys. Rev. Lett.}\ }\textbf
  {\bibinfo {volume} {122}},\ \bibinfo {pages} {010404} (\bibinfo {year}
  {2019})}\BibitemShut {NoStop}%
\bibitem [{\citenamefont {Lang}\ and\ \citenamefont
  {Sch\"utzhold}(2019)}]{Lang2019}%
  \BibitemOpen
  \bibfield  {author} {\bibinfo {author} {\bibfnamefont {S.}~\bibnamefont
  {Lang}}\ and\ \bibinfo {author} {\bibfnamefont {R.}~\bibnamefont
  {Sch\"utzhold}},\ }\bibfield  {title} {\bibinfo {title} {Analog of
  cosmological particle creation in electromagnetic waveguides},\ }\href
  {https://doi.org/10.1103/PhysRevD.100.065003} {\bibfield  {journal} {\bibinfo
   {journal} {Phys. Rev. D}\ }\textbf {\bibinfo {volume} {100}},\ \bibinfo
  {pages} {065003} (\bibinfo {year} {2019})}\BibitemShut {NoStop}%
\bibitem [{\citenamefont {Boiron}\ \emph {et~al.}(2015)\citenamefont {Boiron},
  \citenamefont {Fabbri}, \citenamefont {Larr\'e}, \citenamefont {Pavloff},
  \citenamefont {Westbrook},\ and\ \citenamefont {Zi\ifmmode~\acute{n}\else
  \'{n}\fi{}}}]{Boiron2015}%
  \BibitemOpen
  \bibfield  {author} {\bibinfo {author} {\bibfnamefont {D.}~\bibnamefont
  {Boiron}}, \bibinfo {author} {\bibfnamefont {A.}~\bibnamefont {Fabbri}},
  \bibinfo {author} {\bibfnamefont {P.-E.}\ \bibnamefont {Larr\'e}}, \bibinfo
  {author} {\bibfnamefont {N.}~\bibnamefont {Pavloff}}, \bibinfo {author}
  {\bibfnamefont {C.~I.}\ \bibnamefont {Westbrook}},\ and\ \bibinfo {author}
  {\bibfnamefont {P.}~\bibnamefont {Zi\ifmmode~\acute{n}\else \'{n}\fi{}}},\
  }\bibfield  {title} {\bibinfo {title} {Quantum signature of analog {H}awking
  radiation in momentum space},\ }\href
  {https://doi.org/10.1103/PhysRevLett.115.025301} {\bibfield  {journal}
  {\bibinfo  {journal} {Phys. Rev. Lett.}\ }\textbf {\bibinfo {volume} {115}},\
  \bibinfo {pages} {025301} (\bibinfo {year} {2015})}\BibitemShut {NoStop}%
\bibitem [{\citenamefont {Rold\'an-Molina}\ \emph {et~al.}(2017)\citenamefont
  {Rold\'an-Molina}, \citenamefont {Nunez},\ and\ \citenamefont
  {Duine}}]{Roldan-Molina2017}%
  \BibitemOpen
  \bibfield  {author} {\bibinfo {author} {\bibfnamefont {A.}~\bibnamefont
  {Rold\'an-Molina}}, \bibinfo {author} {\bibfnamefont {A.~S.}\ \bibnamefont
  {Nunez}},\ and\ \bibinfo {author} {\bibfnamefont {R.~A.}\ \bibnamefont
  {Duine}},\ }\bibfield  {title} {\bibinfo {title} {Magnonic black holes},\
  }\href {https://doi.org/10.1103/PhysRevLett.118.061301} {\bibfield  {journal}
  {\bibinfo  {journal} {Phys. Rev. Lett.}\ }\textbf {\bibinfo {volume} {118}},\
  \bibinfo {pages} {061301} (\bibinfo {year} {2017})}\BibitemShut {NoStop}%
\bibitem [{\citenamefont {Chuang}\ \emph {et~al.}(1991)\citenamefont {Chuang},
  \citenamefont {Durrer}, \citenamefont {Turok},\ and\ \citenamefont
  {Yurke}}]{Chuang1991}%
  \BibitemOpen
  \bibfield  {author} {\bibinfo {author} {\bibfnamefont {I.}~\bibnamefont
  {Chuang}}, \bibinfo {author} {\bibfnamefont {R.}~\bibnamefont {Durrer}},
  \bibinfo {author} {\bibfnamefont {N.}~\bibnamefont {Turok}},\ and\ \bibinfo
  {author} {\bibfnamefont {B.}~\bibnamefont {Yurke}},\ }\bibfield  {title}
  {\bibinfo {title} {Cosmology in the laboratory: Defect dynamics in liquid
  crystals},\ }\href {https://doi.org/10.1126/science.251.4999.1336} {\bibfield
   {journal} {\bibinfo  {journal} {Science}\ }\textbf {\bibinfo {volume}
  {251}},\ \bibinfo {pages} {1336} (\bibinfo {year} {1991})}\BibitemShut
  {NoStop}%
\bibitem [{\citenamefont {Barcel{\'o}}\ \emph {et~al.}(2011)\citenamefont
  {Barcel{\'o}}, \citenamefont {Liberati},\ and\ \citenamefont
  {Visser}}]{Barcelo2011}%
  \BibitemOpen
  \bibfield  {author} {\bibinfo {author} {\bibfnamefont {C.}~\bibnamefont
  {Barcel{\'o}}}, \bibinfo {author} {\bibfnamefont {S.}~\bibnamefont
  {Liberati}},\ and\ \bibinfo {author} {\bibfnamefont {M.}~\bibnamefont
  {Visser}},\ }\bibfield  {title} {\bibinfo {title} {Analogue gravity},\ }\href
  {https://link.springer.com/article/10.12942/lrr-2011-3} {\bibfield  {journal}
  {\bibinfo  {journal} {Living Rev. Relativity}\ }\textbf {\bibinfo {volume}
  {14}},\ \bibinfo {pages} {3} (\bibinfo {year} {2011})}\BibitemShut {NoStop}%
\bibitem [{\citenamefont {Faccio}\ \emph {et~al.}(2013)\citenamefont {Faccio},
  \citenamefont {Belgiorno}, \citenamefont {Cacciatori}, \citenamefont
  {Gorini}, \citenamefont {Liberati},\ and\ \citenamefont
  {Moschella}}]{Faccio2013}%
  \BibitemOpen
  \bibfield  {author} {\bibinfo {author} {\bibfnamefont {D.}~\bibnamefont
  {Faccio}}, \bibinfo {author} {\bibfnamefont {F.}~\bibnamefont {Belgiorno}},
  \bibinfo {author} {\bibfnamefont {S.}~\bibnamefont {Cacciatori}}, \bibinfo
  {author} {\bibfnamefont {V.}~\bibnamefont {Gorini}}, \bibinfo {author}
  {\bibfnamefont {S.}~\bibnamefont {Liberati}},\ and\ \bibinfo {author}
  {\bibfnamefont {U.}~\bibnamefont {Moschella}},\ }\href@noop {} {\emph
  {\bibinfo {title} {Analogue gravity phenomenology: analogue spacetimes and
  horizons, from theory to experiment}}},\ Vol.\ \bibinfo {volume} {870}\
  (\bibinfo  {publisher} {Springer},\ \bibinfo {year} {2013})\BibitemShut
  {NoStop}%
\bibitem [{\citenamefont {Alsing}\ \emph {et~al.}(2005)\citenamefont {Alsing},
  \citenamefont {Dowling},\ and\ \citenamefont {Milburn}}]{Alsing2005}%
  \BibitemOpen
  \bibfield  {author} {\bibinfo {author} {\bibfnamefont {P.~M.}\ \bibnamefont
  {Alsing}}, \bibinfo {author} {\bibfnamefont {J.~P.}\ \bibnamefont
  {Dowling}},\ and\ \bibinfo {author} {\bibfnamefont {G.~J.}\ \bibnamefont
  {Milburn}},\ }\bibfield  {title} {\bibinfo {title} {Ion trap simulations of
  quantum fields in an expanding universe},\ }\href
  {https://doi.org/10.1103/PhysRevLett.94.220401} {\bibfield  {journal}
  {\bibinfo  {journal} {Phys. Rev. Lett.}\ }\textbf {\bibinfo {volume} {94}},\
  \bibinfo {pages} {220401} (\bibinfo {year} {2005})}\BibitemShut {NoStop}%
\bibitem [{\citenamefont {Sch\"utzhold}\ \emph {et~al.}(2007)\citenamefont
  {Sch\"utzhold}, \citenamefont {Uhlmann}, \citenamefont {Petersen},
  \citenamefont {Schmitz}, \citenamefont {Friedenauer},\ and\ \citenamefont
  {Sch\"atz}}]{Schutzhold2007}%
  \BibitemOpen
  \bibfield  {author} {\bibinfo {author} {\bibfnamefont {R.}~\bibnamefont
  {Sch\"utzhold}}, \bibinfo {author} {\bibfnamefont {M.}~\bibnamefont
  {Uhlmann}}, \bibinfo {author} {\bibfnamefont {L.}~\bibnamefont {Petersen}},
  \bibinfo {author} {\bibfnamefont {H.}~\bibnamefont {Schmitz}}, \bibinfo
  {author} {\bibfnamefont {A.}~\bibnamefont {Friedenauer}},\ and\ \bibinfo
  {author} {\bibfnamefont {T.}~\bibnamefont {Sch\"atz}},\ }\bibfield  {title}
  {\bibinfo {title} {Analogue of cosmological particle creation in an ion
  trap},\ }\href {https://doi.org/10.1103/PhysRevLett.99.201301} {\bibfield
  {journal} {\bibinfo  {journal} {Phys. Rev. Lett.}\ }\textbf {\bibinfo
  {volume} {99}},\ \bibinfo {pages} {201301} (\bibinfo {year}
  {2007})}\BibitemShut {NoStop}%
\bibitem [{\citenamefont {Menicucci}\ \emph {et~al.}(2010)\citenamefont
  {Menicucci}, \citenamefont {Olson},\ and\ \citenamefont
  {Milburn}}]{Nicolas2010}%
  \BibitemOpen
  \bibfield  {author} {\bibinfo {author} {\bibfnamefont {N.~C.}\ \bibnamefont
  {Menicucci}}, \bibinfo {author} {\bibfnamefont {S.~J.}\ \bibnamefont
  {Olson}},\ and\ \bibinfo {author} {\bibfnamefont {G.~J.}\ \bibnamefont
  {Milburn}},\ }\bibfield  {title} {\bibinfo {title} {Simulating quantum
  effects of cosmological expansion using a static ion trap},\ }\href
  {https://doi.org/10.1088/1367-2630/12/9/095019} {\bibfield  {journal}
  {\bibinfo  {journal} {New Journal of Physics}\ }\textbf {\bibinfo {volume}
  {12}},\ \bibinfo {pages} {095019} (\bibinfo {year} {2010})}\BibitemShut
  {NoStop}%
\bibitem [{\citenamefont {Casanova}\ \emph {et~al.}(2010)\citenamefont
  {Casanova}, \citenamefont {Garc\'{\i}a-Ripoll}, \citenamefont {Gerritsma},
  \citenamefont {Roos},\ and\ \citenamefont {Solano}}]{Casanova2010}%
  \BibitemOpen
  \bibfield  {author} {\bibinfo {author} {\bibfnamefont {J.}~\bibnamefont
  {Casanova}}, \bibinfo {author} {\bibfnamefont {J.~J.}\ \bibnamefont
  {Garc\'{\i}a-Ripoll}}, \bibinfo {author} {\bibfnamefont {R.}~\bibnamefont
  {Gerritsma}}, \bibinfo {author} {\bibfnamefont {C.~F.}\ \bibnamefont
  {Roos}},\ and\ \bibinfo {author} {\bibfnamefont {E.}~\bibnamefont {Solano}},\
  }\bibfield  {title} {\bibinfo {title} {Klein tunneling and {D}irac potentials
  in trapped ions},\ }\href {https://doi.org/10.1103/PhysRevA.82.020101}
  {\bibfield  {journal} {\bibinfo  {journal} {Phys. Rev. A}\ }\textbf {\bibinfo
  {volume} {82}},\ \bibinfo {pages} {020101} (\bibinfo {year}
  {2010})}\BibitemShut {NoStop}%
\bibitem [{\citenamefont {Gerritsma}\ \emph {et~al.}(2011)\citenamefont
  {Gerritsma}, \citenamefont {Lanyon}, \citenamefont {Kirchmair}, \citenamefont
  {Z\"ahringer}, \citenamefont {Hempel}, \citenamefont {Casanova},
  \citenamefont {Garc\'{\i}a-Ripoll}, \citenamefont {Solano}, \citenamefont
  {Blatt},\ and\ \citenamefont {Roos}}]{Gerritsma2011}%
  \BibitemOpen
  \bibfield  {author} {\bibinfo {author} {\bibfnamefont {R.}~\bibnamefont
  {Gerritsma}}, \bibinfo {author} {\bibfnamefont {B.~P.}\ \bibnamefont
  {Lanyon}}, \bibinfo {author} {\bibfnamefont {G.}~\bibnamefont {Kirchmair}},
  \bibinfo {author} {\bibfnamefont {F.}~\bibnamefont {Z\"ahringer}}, \bibinfo
  {author} {\bibfnamefont {C.}~\bibnamefont {Hempel}}, \bibinfo {author}
  {\bibfnamefont {J.}~\bibnamefont {Casanova}}, \bibinfo {author}
  {\bibfnamefont {J.~J.}\ \bibnamefont {Garc\'{\i}a-Ripoll}}, \bibinfo {author}
  {\bibfnamefont {E.}~\bibnamefont {Solano}}, \bibinfo {author} {\bibfnamefont
  {R.}~\bibnamefont {Blatt}},\ and\ \bibinfo {author} {\bibfnamefont {C.~F.}\
  \bibnamefont {Roos}},\ }\bibfield  {title} {\bibinfo {title} {Quantum
  simulation of the {K}lein paradox with trapped ions},\ }\href
  {https://doi.org/10.1103/PhysRevLett.106.060503} {\bibfield  {journal}
  {\bibinfo  {journal} {Phys. Rev. Lett.}\ }\textbf {\bibinfo {volume} {106}},\
  \bibinfo {pages} {060503} (\bibinfo {year} {2011})}\BibitemShut {NoStop}%
\bibitem [{\citenamefont {Wittemer}\ \emph {et~al.}(2019)\citenamefont
  {Wittemer}, \citenamefont {Hakelberg}, \citenamefont {Kiefer}, \citenamefont
  {Schr\"oder}, \citenamefont {Fey}, \citenamefont {Sch\"utzhold},
  \citenamefont {Warring},\ and\ \citenamefont {Schaetz}}]{Wittemer2019}%
  \BibitemOpen
  \bibfield  {author} {\bibinfo {author} {\bibfnamefont {M.}~\bibnamefont
  {Wittemer}}, \bibinfo {author} {\bibfnamefont {F.}~\bibnamefont {Hakelberg}},
  \bibinfo {author} {\bibfnamefont {P.}~\bibnamefont {Kiefer}}, \bibinfo
  {author} {\bibfnamefont {J.-P.}\ \bibnamefont {Schr\"oder}}, \bibinfo
  {author} {\bibfnamefont {C.}~\bibnamefont {Fey}}, \bibinfo {author}
  {\bibfnamefont {R.}~\bibnamefont {Sch\"utzhold}}, \bibinfo {author}
  {\bibfnamefont {U.}~\bibnamefont {Warring}},\ and\ \bibinfo {author}
  {\bibfnamefont {T.}~\bibnamefont {Schaetz}},\ }\bibfield  {title} {\bibinfo
  {title} {Phonon pair creation by inflating quantum fluctuations in an ion
  trap},\ }\href {https://doi.org/10.1103/PhysRevLett.123.180502} {\bibfield
  {journal} {\bibinfo  {journal} {Phys. Rev. Lett.}\ }\textbf {\bibinfo
  {volume} {123}},\ \bibinfo {pages} {180502} (\bibinfo {year}
  {2019})}\BibitemShut {NoStop}%
\bibitem [{\citenamefont {Tian}\ \emph {et~al.}(2020)\citenamefont {Tian},
  \citenamefont {Lin}, \citenamefont {Fischer},\ and\ \citenamefont
  {Du}}]{TianZehua2020}%
  \BibitemOpen
  \bibfield  {author} {\bibinfo {author} {\bibfnamefont {Z.}~\bibnamefont
  {Tian}}, \bibinfo {author} {\bibfnamefont {Y.}~\bibnamefont {Lin}}, \bibinfo
  {author} {\bibfnamefont {U.~R.}\ \bibnamefont {Fischer}},\ and\ \bibinfo
  {author} {\bibfnamefont {J.}~\bibnamefont {Du}},\ }\bibfield  {title}
  {\bibinfo {title} {Verifying the upper bound on the speed of scrambling with
  the analogue {H}awking radiation of trapped ions},\ }\href
  {https://arxiv.org/abs/2007.05949} {\bibfield  {journal} {\bibinfo  {journal}
  {arXiv:2007.05949}\ } (\bibinfo {year} {2020})}\BibitemShut {NoStop}%
\bibitem [{\citenamefont {Yang}\ \emph {et~al.}(2020)\citenamefont {Yang},
  \citenamefont {Liu}, \citenamefont {Zhu}, \citenamefont {Luo},\ and\
  \citenamefont {Cai}}]{YangRunQiu2020}%
  \BibitemOpen
  \bibfield  {author} {\bibinfo {author} {\bibfnamefont {R.-Q.}\ \bibnamefont
  {Yang}}, \bibinfo {author} {\bibfnamefont {H.}~\bibnamefont {Liu}}, \bibinfo
  {author} {\bibfnamefont {S.}~\bibnamefont {Zhu}}, \bibinfo {author}
  {\bibfnamefont {L.}~\bibnamefont {Luo}},\ and\ \bibinfo {author}
  {\bibfnamefont {R.-G.}\ \bibnamefont {Cai}},\ }\bibfield  {title} {\bibinfo
  {title} {Simulating quantum field theory in curved spacetime with quantum
  many-body systems},\ }\href
  {https://doi.org/10.1103/PhysRevResearch.2.023107} {\bibfield  {journal}
  {\bibinfo  {journal} {Phys. Rev. Research}\ }\textbf {\bibinfo {volume}
  {2}},\ \bibinfo {pages} {023107} (\bibinfo {year} {2020})}\BibitemShut
  {NoStop}%
\bibitem [{\citenamefont {Nation}\ \emph {et~al.}(2009)\citenamefont {Nation},
  \citenamefont {Blencowe}, \citenamefont {Rimberg},\ and\ \citenamefont
  {Buks}}]{Nation2009}%
  \BibitemOpen
  \bibfield  {author} {\bibinfo {author} {\bibfnamefont {P.~D.}\ \bibnamefont
  {Nation}}, \bibinfo {author} {\bibfnamefont {M.~P.}\ \bibnamefont
  {Blencowe}}, \bibinfo {author} {\bibfnamefont {A.~J.}\ \bibnamefont
  {Rimberg}},\ and\ \bibinfo {author} {\bibfnamefont {E.}~\bibnamefont
  {Buks}},\ }\bibfield  {title} {\bibinfo {title} {Analogue {H}awking radiation
  in a dc-{SQUID} array transmission line},\ }\href
  {https://doi.org/10.1103/PhysRevLett.103.087004} {\bibfield  {journal}
  {\bibinfo  {journal} {Phys. Rev. Lett.}\ }\textbf {\bibinfo {volume} {103}},\
  \bibinfo {pages} {087004} (\bibinfo {year} {2009})}\BibitemShut {NoStop}%
\bibitem [{\citenamefont {Sab\'{\i}n}(2016)}]{Sabin2016}%
  \BibitemOpen
  \bibfield  {author} {\bibinfo {author} {\bibfnamefont {C.}~\bibnamefont
  {Sab\'{\i}n}},\ }\bibfield  {title} {\bibinfo {title} {Quantum simulation of
  traversable wormhole spacetimes in a dc-{SQUID} array},\ }\href
  {https://doi.org/10.1103/PhysRevD.94.081501} {\bibfield  {journal} {\bibinfo
  {journal} {Phys. Rev. D}\ }\textbf {\bibinfo {volume} {94}},\ \bibinfo
  {pages} {081501} (\bibinfo {year} {2016})}\BibitemShut {NoStop}%
\bibitem [{\citenamefont {Tian}\ \emph {et~al.}(2017)\citenamefont {Tian},
  \citenamefont {Jing},\ and\ \citenamefont {Dragan}}]{TianZehua2017}%
  \BibitemOpen
  \bibfield  {author} {\bibinfo {author} {\bibfnamefont {Z.}~\bibnamefont
  {Tian}}, \bibinfo {author} {\bibfnamefont {J.}~\bibnamefont {Jing}},\ and\
  \bibinfo {author} {\bibfnamefont {A.}~\bibnamefont {Dragan}},\ }\bibfield
  {title} {\bibinfo {title} {Analog cosmological particle generation in a
  superconducting circuit},\ }\href
  {https://doi.org/10.1103/PhysRevD.95.125003} {\bibfield  {journal} {\bibinfo
  {journal} {Phys. Rev. D}\ }\textbf {\bibinfo {volume} {95}},\ \bibinfo
  {pages} {125003} (\bibinfo {year} {2017})}\BibitemShut {NoStop}%
\bibitem [{\citenamefont {Tian}\ and\ \citenamefont
  {Du}(2019)}]{TianZehua2019}%
  \BibitemOpen
  \bibfield  {author} {\bibinfo {author} {\bibfnamefont {Z.}~\bibnamefont
  {Tian}}\ and\ \bibinfo {author} {\bibfnamefont {J.}~\bibnamefont {Du}},\
  }\bibfield  {title} {\bibinfo {title} {Analogue {H}awking radiation and
  quantum soliton evaporation in a superconducting circuit},\ }\href
  {https://link.springer.com/article/10.1140%2Fepjc%2Fs10052-019-7514-9}
  {\bibfield  {journal} {\bibinfo  {journal} {The European Physical Journal C}\
  }\textbf {\bibinfo {volume} {79}},\ \bibinfo {pages} {1} (\bibinfo {year}
  {2019})}\BibitemShut {NoStop}%
\bibitem [{\citenamefont {Collas}\ and\ \citenamefont
  {Klein}(2019)}]{Collas2019}%
  \BibitemOpen
  \bibfield  {author} {\bibinfo {author} {\bibfnamefont {P.}~\bibnamefont
  {Collas}}\ and\ \bibinfo {author} {\bibfnamefont {D.}~\bibnamefont {Klein}},\
  }\href {https://doi.org/10.1007/978-3-030-14825-6} {\emph {\bibinfo {title}
  {The {D}irac equation in curved spacetime}}}\ (\bibinfo  {publisher}
  {Springer International Publishing},\ \bibinfo {year} {2019})\BibitemShut
  {NoStop}%
\bibitem [{\citenamefont {Pedernales}\ \emph {et~al.}(2018)\citenamefont
  {Pedernales}, \citenamefont {Beau}, \citenamefont {Pittman}, \citenamefont
  {Egusquiza}, \citenamefont {Lamata}, \citenamefont {Solano},\ and\
  \citenamefont {del Campo}}]{Pedernales2018}%
  \BibitemOpen
  \bibfield  {author} {\bibinfo {author} {\bibfnamefont {J.~S.}\ \bibnamefont
  {Pedernales}}, \bibinfo {author} {\bibfnamefont {M.}~\bibnamefont {Beau}},
  \bibinfo {author} {\bibfnamefont {S.~M.}\ \bibnamefont {Pittman}}, \bibinfo
  {author} {\bibfnamefont {I.~L.}\ \bibnamefont {Egusquiza}}, \bibinfo {author}
  {\bibfnamefont {L.}~\bibnamefont {Lamata}}, \bibinfo {author} {\bibfnamefont
  {E.}~\bibnamefont {Solano}},\ and\ \bibinfo {author} {\bibfnamefont
  {A.}~\bibnamefont {del Campo}},\ }\bibfield  {title} {\bibinfo {title}
  {{D}irac equation in ($1+1$)-dimensional curved spacetime and the multiphoton
  quantum {R}abi model},\ }\href
  {https://doi.org/10.1103/PhysRevLett.120.160403} {\bibfield  {journal}
  {\bibinfo  {journal} {Phys. Rev. Lett.}\ }\textbf {\bibinfo {volume} {120}},\
  \bibinfo {pages} {160403} (\bibinfo {year} {2018})}\BibitemShut {NoStop}%
\bibitem [{\citenamefont {Garc\'{\i}a}\ and\ \citenamefont
  {Sab\'{\i}n}(2019)}]{Garcia2019}%
  \BibitemOpen
  \bibfield  {author} {\bibinfo {author} {\bibfnamefont {J.~F.}\ \bibnamefont
  {Garc\'{\i}a}}\ and\ \bibinfo {author} {\bibfnamefont {C.}~\bibnamefont
  {Sab\'{\i}n}},\ }\bibfield  {title} {\bibinfo {title} {{D}irac equation in
  exotic spacetimes},\ }\href {https://doi.org/10.1103/PhysRevD.99.025008}
  {\bibfield  {journal} {\bibinfo  {journal} {Phys. Rev. D}\ }\textbf {\bibinfo
  {volume} {99}},\ \bibinfo {pages} {025008} (\bibinfo {year}
  {2019})}\BibitemShut {NoStop}%
\bibitem [{\citenamefont {Lamata}\ \emph {et~al.}(2007)\citenamefont {Lamata},
  \citenamefont {Le\'on}, \citenamefont {Sch\"atz},\ and\ \citenamefont
  {Solano}}]{Lamata2007}%
  \BibitemOpen
  \bibfield  {author} {\bibinfo {author} {\bibfnamefont {L.}~\bibnamefont
  {Lamata}}, \bibinfo {author} {\bibfnamefont {J.}~\bibnamefont {Le\'on}},
  \bibinfo {author} {\bibfnamefont {T.}~\bibnamefont {Sch\"atz}},\ and\
  \bibinfo {author} {\bibfnamefont {E.}~\bibnamefont {Solano}},\ }\bibfield
  {title} {\bibinfo {title} {{D}irac equation and quantum relativistic effects
  in a single trapped ion},\ }\href
  {https://doi.org/10.1103/PhysRevLett.98.253005} {\bibfield  {journal}
  {\bibinfo  {journal} {Phys. Rev. Lett.}\ }\textbf {\bibinfo {volume} {98}},\
  \bibinfo {pages} {253005} (\bibinfo {year} {2007})}\BibitemShut {NoStop}%
\bibitem [{\citenamefont {Alcubierre}(1994)}]{alcubierre1994}%
  \BibitemOpen
  \bibfield  {author} {\bibinfo {author} {\bibfnamefont {M.}~\bibnamefont
  {Alcubierre}},\ }\bibfield  {title} {\bibinfo {title} {The warp drive:
  hyper-fast travel within general relativity},\ }\href
  {https://doi.org/10.1088/0264-9381/11/5/001} {\bibfield  {journal} {\bibinfo
  {journal} {Classical and Quantum Gravity}\ }\textbf {\bibinfo {volume}
  {11}},\ \bibinfo {pages} {L73} (\bibinfo {year} {1994})}\BibitemShut
  {NoStop}%
\bibitem [{\citenamefont {Finazzi}\ \emph {et~al.}(2009)\citenamefont
  {Finazzi}, \citenamefont {Liberati},\ and\ \citenamefont
  {Barcel\'o}}]{Finazzi2009}%
  \BibitemOpen
  \bibfield  {author} {\bibinfo {author} {\bibfnamefont {S.}~\bibnamefont
  {Finazzi}}, \bibinfo {author} {\bibfnamefont {S.}~\bibnamefont {Liberati}},\
  and\ \bibinfo {author} {\bibfnamefont {C.}~\bibnamefont {Barcel\'o}},\
  }\bibfield  {title} {\bibinfo {title} {Semiclassical instability of dynamical
  warp drives},\ }\href {https://doi.org/10.1103/PhysRevD.79.124017} {\bibfield
   {journal} {\bibinfo  {journal} {Phys. Rev. D}\ }\textbf {\bibinfo {volume}
  {79}},\ \bibinfo {pages} {124017} (\bibinfo {year} {2009})}\BibitemShut
  {NoStop}%
\bibitem [{\citenamefont {White}(2003)}]{White2003}%
  \BibitemOpen
  \bibfield  {author} {\bibinfo {author} {\bibfnamefont {H.}~\bibnamefont
  {White}},\ }\bibfield  {title} {\bibinfo {title} {A discussion of space-time
  metric engineering},\ }\href
  {https://link.springer.com/article/10.1023/A:1026247026218} {\bibfield
  {journal} {\bibinfo  {journal} {Gen. Relativ. Gravit.}\ }\textbf {\bibinfo
  {volume} {35}},\ \bibinfo {pages} {2025} (\bibinfo {year}
  {2003})}\BibitemShut {NoStop}%
\bibitem [{\citenamefont {White}(2011)}]{White2011}%
  \BibitemOpen
  \bibfield  {author} {\bibinfo {author} {\bibfnamefont {H.}~\bibnamefont
  {White}},\ }\bibfield  {title} {\bibinfo {title} {Warp field mechanics 101},\
  }\href
  {https://ntrs.nasa.gov/api/citations/20110015936/downloads/20110015936.pdf}
  {\bibfield  {journal} {\bibinfo  {journal} {J. Br. Interplanet. Soc.}\
  }\textbf {\bibinfo {volume} {66}},\ \bibinfo {pages} {242} (\bibinfo {year}
  {2011})}\BibitemShut {NoStop}%
\bibitem [{\citenamefont {Gonz\'alez-D\'{\i}az}(2000)}]{Gonzalez-Diaz2000}%
  \BibitemOpen
  \bibfield  {author} {\bibinfo {author} {\bibfnamefont {P.~F.}\ \bibnamefont
  {Gonz\'alez-D\'{\i}az}},\ }\bibfield  {title} {\bibinfo {title} {Warp drive
  space-time},\ }\href {https://doi.org/10.1103/PhysRevD.62.044005} {\bibfield
  {journal} {\bibinfo  {journal} {Phys. Rev. D}\ }\textbf {\bibinfo {volume}
  {62}},\ \bibinfo {pages} {044005} (\bibinfo {year} {2000})}\BibitemShut
  {NoStop}%
\bibitem [{\citenamefont {Mann}\ \emph {et~al.}(1991)\citenamefont {Mann},
  \citenamefont {Morsink}, \citenamefont {Sikkema},\ and\ \citenamefont
  {Steele}}]{Mann1991}%
  \BibitemOpen
  \bibfield  {author} {\bibinfo {author} {\bibfnamefont {R.~B.}\ \bibnamefont
  {Mann}}, \bibinfo {author} {\bibfnamefont {S.~M.}\ \bibnamefont {Morsink}},
  \bibinfo {author} {\bibfnamefont {A.~E.}\ \bibnamefont {Sikkema}},\ and\
  \bibinfo {author} {\bibfnamefont {T.~G.}\ \bibnamefont {Steele}},\ }\bibfield
   {title} {\bibinfo {title} {Semiclassical gravity in 1+1 dimensions},\ }\href
  {https://doi.org/10.1103/PhysRevD.43.3948} {\bibfield  {journal} {\bibinfo
  {journal} {Phys. Rev. D}\ }\textbf {\bibinfo {volume} {43}},\ \bibinfo
  {pages} {3948} (\bibinfo {year} {1991})}\BibitemShut {NoStop}%
\bibitem [{Sup()}]{Supplement}%
  \BibitemOpen
  \href@noop {} {\bibinfo  {journal} {Further details of analysis and
  calculation areare available as supplementary material, which includes
  Refs.~\cite{Mann1991,Collas2019,Burd2021,Sorensen1999,Srinivas2019}}\
  }\BibitemShut {NoStop}%
\bibitem [{\citenamefont {Thaller}(2013)}]{thaller2013}%
  \BibitemOpen
\bibfield  {journal} {  }\bibfield  {author} {\bibinfo {author} {\bibfnamefont
  {B.}~\bibnamefont {Thaller}},\ }\href
  {https://www.springer.com/gp/book/9783540548836} {\emph {\bibinfo {title}
  {The {D}irac equation}}}\ (\bibinfo  {publisher} {Springer Science \&
  Business Media},\ \bibinfo {year} {2013})\BibitemShut {NoStop}%
\bibitem [{\citenamefont {Burd}\ \emph {et~al.}(2021)\citenamefont {Burd},
  \citenamefont {Srinivas}, \citenamefont {Knaack}, \citenamefont {Ge},
  \citenamefont {Wilson}, \citenamefont {Wineland}, \citenamefont {Leibfried},
  \citenamefont {Bollinger}, \citenamefont {Allcock},\ and\ \citenamefont
  {Slichter}}]{Burd2021}%
  \BibitemOpen
  \bibfield  {author} {\bibinfo {author} {\bibfnamefont {S.~C.}\ \bibnamefont
  {Burd}}, \bibinfo {author} {\bibfnamefont {R.}~\bibnamefont {Srinivas}},
  \bibinfo {author} {\bibfnamefont {H.~M.}\ \bibnamefont {Knaack}}, \bibinfo
  {author} {\bibfnamefont {W.}~\bibnamefont {Ge}}, \bibinfo {author}
  {\bibfnamefont {A.~C.}\ \bibnamefont {Wilson}}, \bibinfo {author}
  {\bibfnamefont {D.~J.}\ \bibnamefont {Wineland}}, \bibinfo {author}
  {\bibfnamefont {D.}~\bibnamefont {Leibfried}}, \bibinfo {author}
  {\bibfnamefont {J.~J.}\ \bibnamefont {Bollinger}}, \bibinfo {author}
  {\bibfnamefont {D.~T.~C.}\ \bibnamefont {Allcock}},\ and\ \bibinfo {author}
  {\bibfnamefont {D.~H.}\ \bibnamefont {Slichter}},\ }\bibfield  {title}
  {\bibinfo {title} {Quantum amplification of boson-mediated interactions},\
  }\bibfield  {journal} {\bibinfo  {journal} {Nature Physics}\ }\href
  {https://doi.org/10.1038/s41567-021-01237-9} {10.1038/s41567-021-01237-9}
  (\bibinfo {year} {2021})\BibitemShut {NoStop}%
\bibitem [{\citenamefont {Burd}\ \emph {et~al.}(2019)\citenamefont {Burd},
  \citenamefont {Srinivas}, \citenamefont {Bollinger}, \citenamefont {Wilson},
  \citenamefont {Wineland}, \citenamefont {Leibfried}, \citenamefont
  {Slichter},\ and\ \citenamefont {Allcock}}]{Burd2019}%
  \BibitemOpen
  \bibfield  {author} {\bibinfo {author} {\bibfnamefont {S.~C.}\ \bibnamefont
  {Burd}}, \bibinfo {author} {\bibfnamefont {R.}~\bibnamefont {Srinivas}},
  \bibinfo {author} {\bibfnamefont {J.~J.}\ \bibnamefont {Bollinger}}, \bibinfo
  {author} {\bibfnamefont {A.~C.}\ \bibnamefont {Wilson}}, \bibinfo {author}
  {\bibfnamefont {D.~J.}\ \bibnamefont {Wineland}}, \bibinfo {author}
  {\bibfnamefont {D.}~\bibnamefont {Leibfried}}, \bibinfo {author}
  {\bibfnamefont {D.~H.}\ \bibnamefont {Slichter}},\ and\ \bibinfo {author}
  {\bibfnamefont {D.~T.~C.}\ \bibnamefont {Allcock}},\ }\bibfield  {title}
  {\bibinfo {title} {Quantum amplification of mechanical oscillator motion},\
  }\href {https://doi.org/10.1126/science.aaw2884} {\bibfield  {journal}
  {\bibinfo  {journal} {Science}\ }\textbf {\bibinfo {volume} {364}},\ \bibinfo
  {pages} {1163} (\bibinfo {year} {2019})}\BibitemShut {NoStop}%
\bibitem [{\citenamefont {S\o{}rensen}\ and\ \citenamefont
  {M\o{}lmer}(1999)}]{Sorensen1999}%
  \BibitemOpen
  \bibfield  {author} {\bibinfo {author} {\bibfnamefont {A.}~\bibnamefont
  {S\o{}rensen}}\ and\ \bibinfo {author} {\bibfnamefont {K.}~\bibnamefont
  {M\o{}lmer}},\ }\bibfield  {title} {\bibinfo {title} {Quantum computation
  with ions in thermal motion},\ }\href
  {https://doi.org/10.1103/PhysRevLett.82.1971} {\bibfield  {journal} {\bibinfo
   {journal} {Phys. Rev. Lett.}\ }\textbf {\bibinfo {volume} {82}},\ \bibinfo
  {pages} {1971} (\bibinfo {year} {1999})}\BibitemShut {NoStop}%
\bibitem [{\citenamefont {Srinivas}\ \emph {et~al.}(2019)\citenamefont
  {Srinivas}, \citenamefont {Burd}, \citenamefont {Sutherland}, \citenamefont
  {Wilson}, \citenamefont {Wineland}, \citenamefont {Leibfried}, \citenamefont
  {Allcock},\ and\ \citenamefont {Slichter}}]{Srinivas2019}%
  \BibitemOpen
  \bibfield  {author} {\bibinfo {author} {\bibfnamefont {R.}~\bibnamefont
  {Srinivas}}, \bibinfo {author} {\bibfnamefont {S.~C.}\ \bibnamefont {Burd}},
  \bibinfo {author} {\bibfnamefont {R.~T.}\ \bibnamefont {Sutherland}},
  \bibinfo {author} {\bibfnamefont {A.~C.}\ \bibnamefont {Wilson}}, \bibinfo
  {author} {\bibfnamefont {D.~J.}\ \bibnamefont {Wineland}}, \bibinfo {author}
  {\bibfnamefont {D.}~\bibnamefont {Leibfried}}, \bibinfo {author}
  {\bibfnamefont {D.~T.~C.}\ \bibnamefont {Allcock}},\ and\ \bibinfo {author}
  {\bibfnamefont {D.~H.}\ \bibnamefont {Slichter}},\ }\bibfield  {title}
  {\bibinfo {title} {Trapped-ion spin-motion coupling with microwaves and a
  near-motional oscillating magnetic field gradient},\ }\href
  {https://doi.org/10.1103/PhysRevLett.122.163201} {\bibfield  {journal}
  {\bibinfo  {journal} {Phys. Rev. Lett.}\ }\textbf {\bibinfo {volume} {122}},\
  \bibinfo {pages} {163201} (\bibinfo {year} {2019})}\BibitemShut {NoStop}%
\end{thebibliography}%
\end{document}